\documentclass[twocolumn,amsmath,amssymb,pra]{revtex4-1}

\usepackage{mathpazo} % add possibly `sc` and `osf` options

\usepackage{graphicx}% Include figure files
\usepackage{dcolumn}% Align table columns on decimal point
\usepackage{bm}% bold math
\usepackage{microtype}
\usepackage{helvet}	

\usepackage[colorlinks=true,allcolors=blue]{hyperref}

\newenvironment{relatedworkarxiv}[6]{\noindent \large \href{http://arxiv.org/abs/#2}{#1} \newline  \vspace*{0.5cm}}

\newenvironment{relatedwork}[6]{\noindent \large \href{http://dx.doi.org/#2}{#1} \newline \textit{#3} \textbf{#4}, #5 (#6)\vspace*{0.5cm}}

\newenvironment{arxiv}[1]{arXiv:\href{http://arxiv.org/abs/#1}{#1}}

\usepackage{caption}
\DeclareCaptionLabelFormat{plain}{Figure #2 $\boldsymbol |$ }

\begin{document}

\title{Discrete Anderson Speckle}

\author{H. Esat Kondakci}\email{esat@creol.ucf.edu}
\affiliation{CREOL, The College of Optics $\&$ Photonics, University of Central Florida, Orlando, FL 32816, USA}

\author{Ayman F. Abouraddy}
\affiliation{CREOL, The College of Optics $\&$ Photonics, University of Central Florida, Orlando, FL 32816, USA}

\author{Bahaa E. A. Saleh}
\affiliation{CREOL, The College of Optics $\&$ Photonics, University of Central Florida, Orlando, FL 32816, USA}

%\date{\today}

\begin{abstract}
When a disordered array of coupled waveguides is illuminated with an extended coherent optical field, discrete speckle develops: partially coherent light with a granular intensity distribution on the lattice sites. The same paradigm applies to a variety of other settings in photonics, such as imperfectly coupled resonators or fibers with randomly coupled cores. Through numerical simulations and analytical modeling, we uncover a set of surprising features that characterize discrete speckle in one- and two-dimensional lattices known to exhibit transverse Anderson localization. Firstly, the fingerprint of localization is embedded in the fluctuations of the discrete speckle and is revealed in the narrowing of the spatial coherence function. Secondly, the transverse coherence length (or speckle grain size) is frozen during propagation. Thirdly, the axial coherence depth is independent of the axial position, thereby resulting in a coherence voxel of fixed volume independently of position. We take these unique features collectively to define a distinct regime that we call discrete Anderson speckle.
\end{abstract}

\small
\maketitle

\begin{figure*}[t!]
	\centering \includegraphics[scale=1]{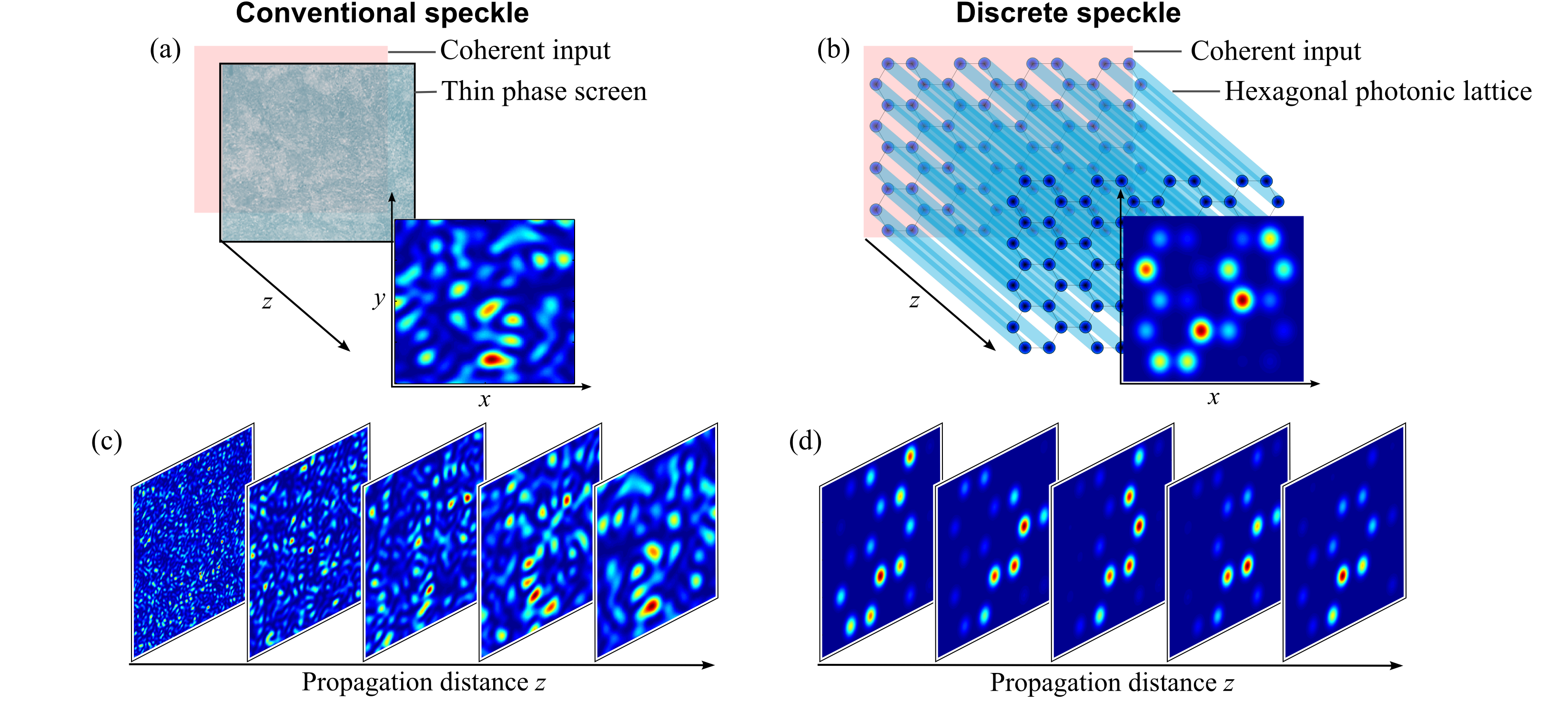} \vspace{-0.2cm}
	\caption{\label{Figure1} \textbf{Conventional speckle and discrete Anderson speckle.} \\ (a) A thin random phase screen ($\sigma_{\phi}\!=\!4\pi$) illuminated with a uniform coherent beam produces conventional speckle. (b) Discrete Anderson speckle is produced from a highly disordered hexagonal (honeycomb) waveguide lattice with maximal off-diagonal disorder when illuminated with a uniform coherent beam. (c) The grain size of conventional speckle increases with propagation distance $z$, while that of (d) discrete Anderson speckle does not.}\vspace{-0.5cm}
\end{figure*}

%\section{Introduction}
Speckle, the granular spatial intensity pattern imbued to a coherent optical field after traversing a disordered medium or reflecting from a rough surface, has been studied for decades extending back to the invention of the laser \cite{Rigden1962, Oliver1963} -- and was known even earlier in radio waves \cite{Ratcliffe1933, Booker1950}. It is a universal phenomenon associated with the interference of random waves. An archetypical arrangement is shown in Fig.~\ref{Figure1}(a) where a coherent wave traverses a thin phase screen and the random phase is converted into random intensity upon free-space propagation, which we refer to hereon as \textit{conventional speckle}. Indeed, the propagation of light in random media or scattering from rough surfaces is critical to practical applications in bio-imaging \cite{Bertolotti2012}, subsurface exploration \cite{Saleh2}, and astronomical observations through turbulent atmospheres \cite{Labeyrie1970}. As such, the study of speckle has recently become of central importance in extracting information from -- or transmitting it through -- complex turbid media \cite{Mosk2012, Redding2012, Katz2012, Katz2014, Matthews2014, Katz2014a, Zhou2014}.

In a multiplicity of contexts, light may be confined to propagate on the sites of a discrete lattice, such as those defined by coupled photorefractive \cite{Schwartz2007a}, semiconductor \cite{Lahini2008a}, or fs laser written silica \cite{Martin2011a} waveguide arrays, random fiber cores \cite{Karbasi2014}, coupled optical resonators \cite{Mookherjea2008} or photonic-crystal waveguides \cite{Topolancik2007}. Whether classical \cite{Schwartz2007a, Lahini2008a, Martin2011a, Karbasi2014, Mookherjea2008, Topolancik2007} or quantum light \cite{Abouraddy2012b, DiGiuseppe2013, Crespi2013,Svozilik2014} is utilized, propagation of an extended coherent field along a disordered photonic lattice produces \textit{discrete speckle} on the lattice sites [Fig.~\ref{Figure1}(b)] -- in contrast to conventional \textit{continuous} speckle. One feature arising from the interference between randomly scattered waves in an otherwise periodic potential is Anderson localization \cite{Anderson1958a, Lagendijk2009a}, which is manifested in the lack of diffusion of the wave function. Optics has enabled direct observation of so-called \textit{transverse} localization \cite{Raedt1989} in coupled waveguide arrays on a transversely disordered lattice \cite{Schwartz2007a, Lahini2008a, Martin2011a, Karbasi2014, Segev2013}, among other realizations \cite{Christodoulides2003a}. Usually in such experiments, only a \textit{single waveguide} is excited and \textit{spatially non-stationary} discrete speckle develops. The typical measure of localization in this scenario is the spatial width of the ensemble-averaged \textit{intensity} distribution of transmitted light \cite{Segev2013}. If instead the waveguides are illuminated by \textit{extended coherent light}, a configuration that has not been thoroughly investigated heretofore  \cite{Lahini2008a,Capeta2011}, a discrete speckle pattern with \textit{spatially invariant statistics} develops that apparently masks the localization signature.

In this paper, we investigate numerically and analytically the statistical properties of discrete speckle in one- and two-dimensional (1D and 2D) disordered Anderson lattices upon extended illumination [Fig.~\ref{Figure1}(b)]. We show that the fingerprint of \textit{localization} is embedded in the \textit{fluctuations} of the emerging light and is thus revealed in the coherence function. We uncover a surprising phenomenon: \textit{the transverse coherence width associated with an extended coherent field is determined by the localization length resulting from a single-site excitation}. Consequently, beyond a critical distance, the transverse speckle grain size 'freezes' upon subsequent propagation along the lattice [Fig.~\ref{Figure1}(d)]. Furthermore, the axial coherence depth is independent of axial position, leading to a coherence `voxel' of fixed volume independent of position. We take these features collectively to define a new regime that we call `discrete Anderson speckle'. Our findings are in contradistinction to the familiar characteristics of conventional speckle \cite{Goodman2007}, wherein the transverse coherence length grows with the free-space propagation distance [Fig.~\ref{Figure1}(c)], as dictated by the van Cittert-Zernike theorem \cite{Goodman2000}.

These findings have their foundation in the different beam propagation dynamics that distinguish discrete lattices from continuous media. Nevertheless, despite the distinctions between conventional and discrete Anderson speckle, both phenomena have a common feature: each system contains a \textit{single} realization of a random function of the transverse coordinate. In conventional speckle the randomness is confined to the thin screen, while in discrete Anderson speckle it extends axially without change. Our results help elucidate the ultimate resolution limits of imaging through an Anderson lattice \cite{Karbasi2014}, introduce new strategies for engineering the spatial optical coherence of a beam of light \cite{Redding2011}, and indicate the potential for tuning higher-order field statistics beyond the Gaussian limits.

Previous investigations of electromagnetic-wave propagation through random media have studied the dimensionless conductance, which is proportional to the transmittance \cite{Imry1999a, Chabanov2000a, Wang2011a}. In such systems, disorder -- and hence localization -- is primarily \textit{axial} instead of transverse. In case of the 1D and 2D photonic systems examined here, the situation is quite distinct since the disorder is \textit{transverse} and back-scattering is not allowed, so that the transmittance is always unity (in the absence of absorption) and the localization is observed in a plane transverse to the propagation axis.

\section{Discrete Optical Lattice Model}
\begin{figure}[b!]
	\centering \vspace{-0.2cm} \includegraphics[scale=1]{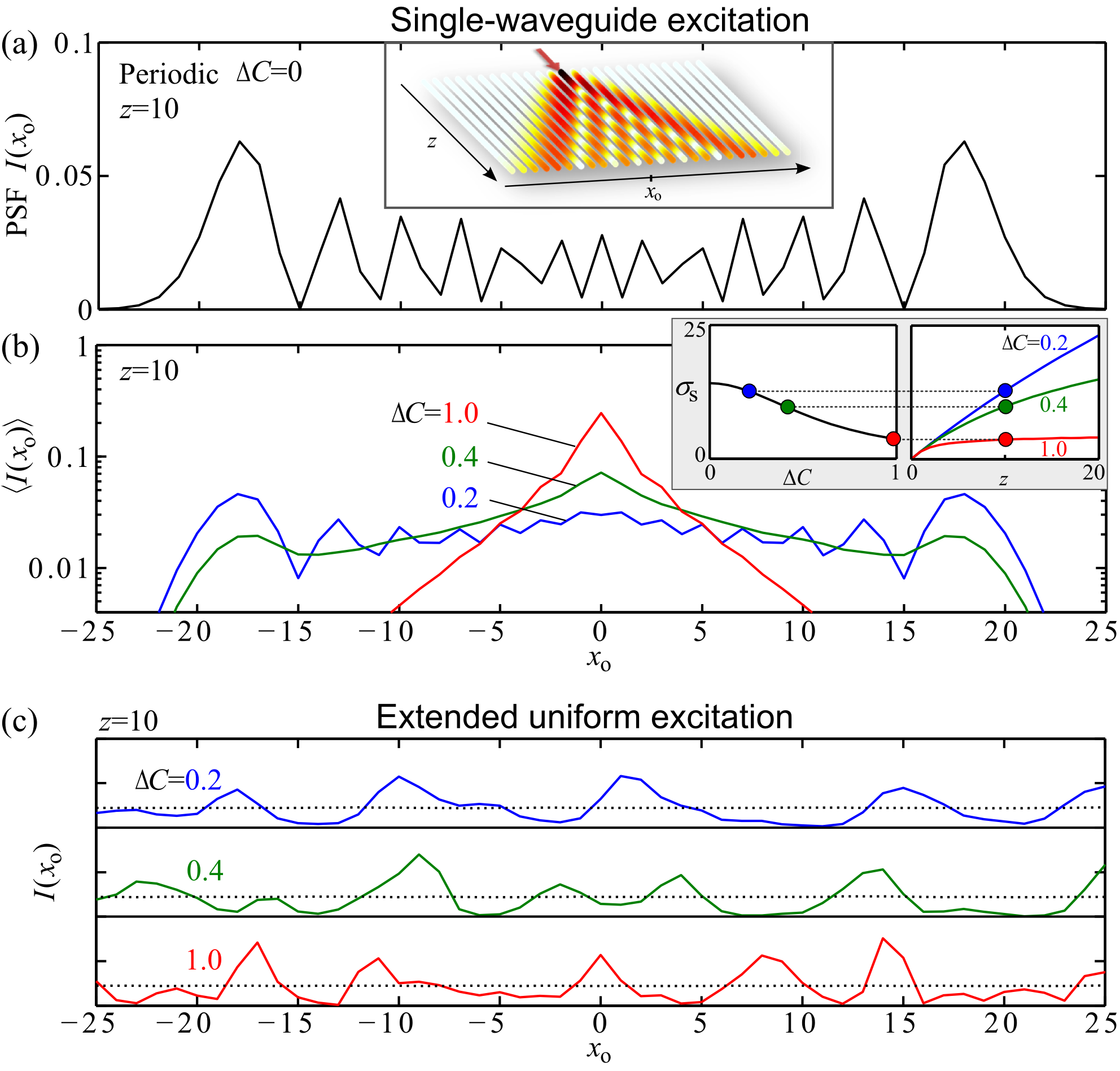} \vspace{-0.2cm}
	\caption{\label{Figure2}  \textbf{Anderson localization and discrete speckle in 1D waveguide lattices.} \\ (a) PSF $I(x_{\mathrm{o}})\!=\!|h(x_{\mathrm{o}},0)|^2$ at $z\!=\!10$ for a 1D \textit{periodic} array for single-waveguide excitation at $x_{\mathrm{i}}\!=\!0$. Inset is a schematic of the configuration. (b) Mean PSF $\langle I(x_{\mathrm{o}})\rangle\!=\!\langle|h(x_{\mathrm{o}},0)|^2\rangle$ for disordered 1D arrays. Insets show the localization length $\sigma_{\mathrm{s}}$ as a function of $\Delta C$ (for fixed $z\!=\!10$) and of $z$ (for fixed values of $\Delta C$). For the values of $\sigma_{\mathrm{s}}$ in the insets, 21 points for $\Delta C$ and 200 for $z$ are chosen. (c) Realizations of discrete speckle at various disorder levels ($z\!=\!10$) for extended uniform coherent input light. The dotted lines are ensemble averages. We use $N_{\mathrm{t}}=151$ throughout.}\vspace{-0.5cm}
\end{figure}

\begin{figure*}[t!]
	\centering \includegraphics[scale=1.1]{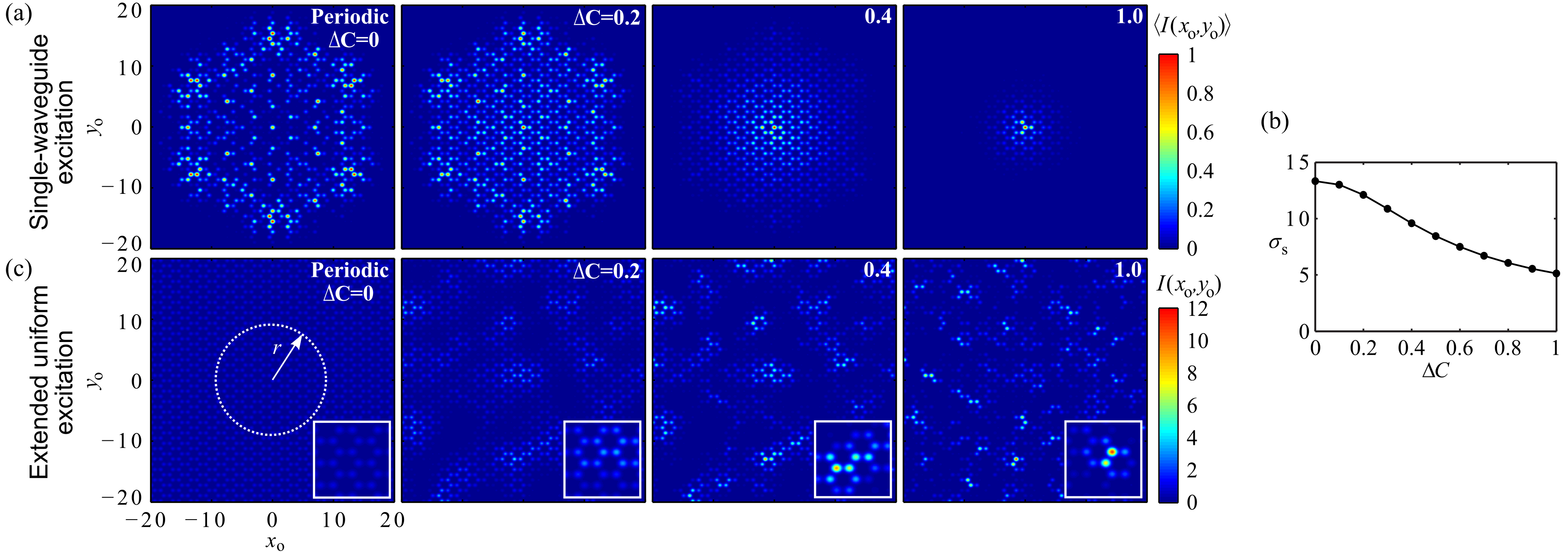} \vspace{-0.2cm}
	\caption{\label{Figure3}  \textbf{Anderson localization and discrete speckle in 2D waveguide lattices.} \\ (a) Mean PSF $\langle I(x_{\mathrm{o}},y_{\mathrm{o}})\rangle =\langle|h(x_{\mathrm{o}},0;y_{\mathrm{o}},0)|^2 \rangle$ for 2D hexagonal (honeycomb) arrays with increasing disorder (from left to right) at $z\!=\!10$. For clarity, each panel is normalized separately and convolved with a gaussian function of width $1.6$ for better visualization. (b) Localization radius $\sigma_{\mathrm{s}}$ as a function of disorder level $\Delta C$. For the values of $\sigma_{\mathrm{s}}$ shown, 11 points for $\Delta C$ are chosen. (c) Individual realizations of discrete speckle at various disorder levels corresponding to the panels in (a) upon extended uniform coherent illumination. Note that the speckle grain size \textit{decreases} with increasing disorder. Insets are magnified by a factor of 2. Speckle contrasts are 0, 0.76, 1.22, 1.24 from left to right. We use $N_{\mathrm{t}}\!\times\!N_{\mathrm{t}}\!=\!101\!\times\!101$ throughout.}\vspace{-0.5cm}
\end{figure*}

Field propagation along a 1D lattice of parallel waveguides with evanescent nearest-neighbor-only coupling [Fig.~\ref{Figure2}] is given by the coupled equations \cite{Christodoulides2003a} 
\begin{equation}\label{eq:general}
i\frac{dE_{x}(z)}{dz}+\beta_{x} E_{x}+C_{x,x-1} E_{x-1}+C_{x,x+1} E_{x+1}=0,
\end{equation}
where $E_{x}(z)$ is the complex optical field in the ${x}^{\text{th}}$ waveguide $({x}\!=\!-N,\ldots,N)$ at axial position $z$, $\beta_{x}$ is the propagation constant of waveguide $x$, and $C_{x,x+1}$ is the coupling coefficient between adjacent waveguides $x$ and $x+1$. The evolution of the input field $E_{\mathrm{i}}(x_{\mathrm{i}})$ to the output $E_{\mathrm{o}}(x_{\mathrm{o}})$ at $z$ may be written as
$\label{eq:field} E_{\mathrm{o}}(x_{\mathrm{o}})\!=\!\sum_{x_{\mathrm{i}}} h(x_{\mathrm{o}},x_{\mathrm{i}})E_{\mathrm{i}}(x_{\mathrm{i}}),$ where $h(x_{\mathrm{o}},x_{\mathrm{i}})$ represents the system's impulse response function after propagating an axial distance $z$ (see the Supplement). The point spread function (PSF) $|h(x_{\mathrm{o}},x_{\mathrm{i}})|^2$ is the corresponding output intensity. This formulation may be readily extended to 2D lattices [Fig.~\ref{Figure3}].

\subsection*{Disorder Classes}
We consider two classes of disorder. The first, \textit{diagonal} disorder \cite{Anderson1958a}, is characterized by constant $C_{x,x+1}=\overline{C}$ and random $\beta_x$ having a uniform probability distribution of mean $\overline{\beta}$ and half width $\Delta\beta$. The second class, \textit{off-diagonal} disorder \cite{Soukoulis1981a}, is characterized by fixed $\beta_x=\overline{\beta}$ and random $C_{x,x+1}$ having a uniform probability distribution of mean $\overline{C}$ and half width $\Delta C$. Both disorder classes exhibit similar behavior in our investigations; we thus report here results for off-diagonal disorder and relegate those for diagonal disorder to the Supplement. The findings of this study are presented in terms of dimensionless variables by writing the coupling coefficients in units of their average $\overline{C}$, and the distance $z$ in units of the coupling length $\ell\!=\!1/\overline{C}$. Throughout, ${\Delta C}$ ranges from 0 to 1. Lattice sizes are chosen large enough so that all the central results in this paper are \textit{independent} of lattice size $N_{\mathrm{t}}=2N+1$. Further details are provided in the Supplement.

\section{Discrete Anderson Speckle:  Transverse Coherence}
\subsection{Anderson Localization}
To set the stage for examining transverse coherence of discrete speckle in Anderson lattices upon uniform illumination, we first describe briefly the results of \textit{single}-waveguide excitation. When disorder is absent ($\Delta C\!=\!0$), ballistic spread leads to an extended output state [Fig.~\ref{Figure2}(a)]. Progressively introducing disorder into the lattice results in a gradual transition to an exponentially localized state [Fig.~\ref{Figure2}(b)] manifested in the pronounced confinement of the mean PSF $\langle|h(x_{\mathrm{o}},0)|^2 \rangle$ around the excitation waveguide, where $\langle\cdot\rangle$ is the ensemble average. In general, similar behavior is observed in 2D lattices [Fig.~\ref{Figure3}(a)]. We define the localization length $\sigma_\mathrm{s}$ as the root-mean-square width of the mean PSF. As shown in the insets of Fig.~\ref{Figure2}(b) and in Fig.~\ref{Figure3}(b), $\sigma_{\mathrm{s}}$ decreases monotonically with increasing $\Delta C$ at fixed distance $z$ in 1D and 2D lattices. On the other hand, $\sigma_{\mathrm{s}}$ typically increases with $z$ at fixed $\Delta C$ until it saturates, a signature of localization, which happens earlier for large $\Delta C$ [Fig.~\ref{Figure2}(b), inset]. For later reference, we note that for short propagation distances at intermediate disorder levels, features of both localized and ballistic states coexist.

\subsection{Transverse Coherence}
We now move on to our investigation of the global statistics of light in Anderson lattices by examining the case of coherent extended uniform illumination. For a 1D array, $E_{\mathrm{i}}(x_{\mathrm{i}})\!=\!1$ and the output field is $E_{\mathrm{o}}(x_{\mathrm{o}})\!=\!\sum_{x_{\mathrm{i}}} h(x_{\mathrm{o}},x_{\mathrm{i}})$, which is a random function of $x_{\mathrm{o}}$ in the case of a disordered lattice; a similar relation holds for 2D arrays. In the absence of disorder, the extended intensity distribution is invariant with respect to propagation [Fig.~\ref{Figure3}(c) for 2D]. Upon introducing disorder, this uniform distribution transitions into a granular intensity pattern $I(x_{\mathrm{o}})\!=\!\langle|E_{\mathrm{o}}(x_{\mathrm{o}})|^{2}\rangle$ defined on the lattice sites -- which we call \textit{discrete} speckle. Examples of individual realizations for 1D and 2D lattices are shown in Fig.~\ref{Figure2}(c) and Fig.~\ref{Figure3}(c), respectively. Several characteristics are immediately apparent in these results.  First, with increasing disorder, the grain size -- which is related to the transverse spatial coherence width -- decreases. On the other hand, the speckle contrast $c$ -- defined as the ratio of the standard deviation in the speckle intensity $\sigma_{\mathrm{I}}$ to its mean intensity $\overline{I}$, $c\!=\!\sigma_{\mathrm{I}}/\overline{I}$ -- increases with disorder. These observations are tell-tale signs of a decrease in the transverse coherence width with increasing disorder. Indeed, these characteristics are shared with conventional speckle \cite{Goodman2007}.

Despite the spatially varying intensity distribution $|E_{\mathrm{o}}(x_{\mathrm{o}})|^{2}$ in the individual realizations for extended input, the statistical homogeneity of this discrete speckle is clear in the uniform distribution obtained upon averaging multiple realizations $\langle|E_{\mathrm{o}}(x_{\mathrm{o}})|^{2}\rangle$ [the dotted lines in Fig.~\ref{Figure2}(c)]. The coherence function at a pair of positions $x_{\mathrm{o}}$ and $x_{\mathrm{o}}+x$ in 1D is therefore a function of only the separation $x$,
\begin{equation}\label{CoherenceFunctionDefinition}
G^{(1)}\!(x_{\mathrm{o}},x_{\mathrm{o}}\!+\!x)\!\!=\!\!G^{(1)}\!(0,x)\!=\!\langle E_{\mathrm{o}}^*(0)E_{\mathrm{o}}(x)\!\rangle\!\!=\!\!\!\sum_{x',x''}\!\!\langle h^{*}(0,x')h(x,x''\!)\!\rangle\!.
\end{equation}
Its normalized version is the complex degree of coherence $g^{(1)}\!(x)\!=\!G^{(1)}\!(0,x)\Big/ \!\!\sqrt{\!G^{(1)}\!(0,0)G^{(1)}\!(x,x)}$, with $0\!\le\!|g^{(1)}\!(x)|\!\le\!1$. In 2D discrete speckle, we similarly write the complex degree of coherence $g^{(1)}(r)$ as a function of the radial separation distance $r$ shown in Fig.~\ref{Figure3}(c). For later reference (see Section 5, `Analytical Model'), we note that transverse spatial invariance results in the double summation in Eq.~\ref{CoherenceFunctionDefinition} separating over the two impulse response functions, such that $G^{(1)}\!(0,x)\!=\!\!\langle\eta\sum_{x''}h(x,x''\!)\!\rangle$, where $\eta\!=\!\sum_{x'}\!h^{*}(0,x')$ is a zero-mean, complex random variable.

\begin{figure}[t!]
	\centering\includegraphics[scale=1]{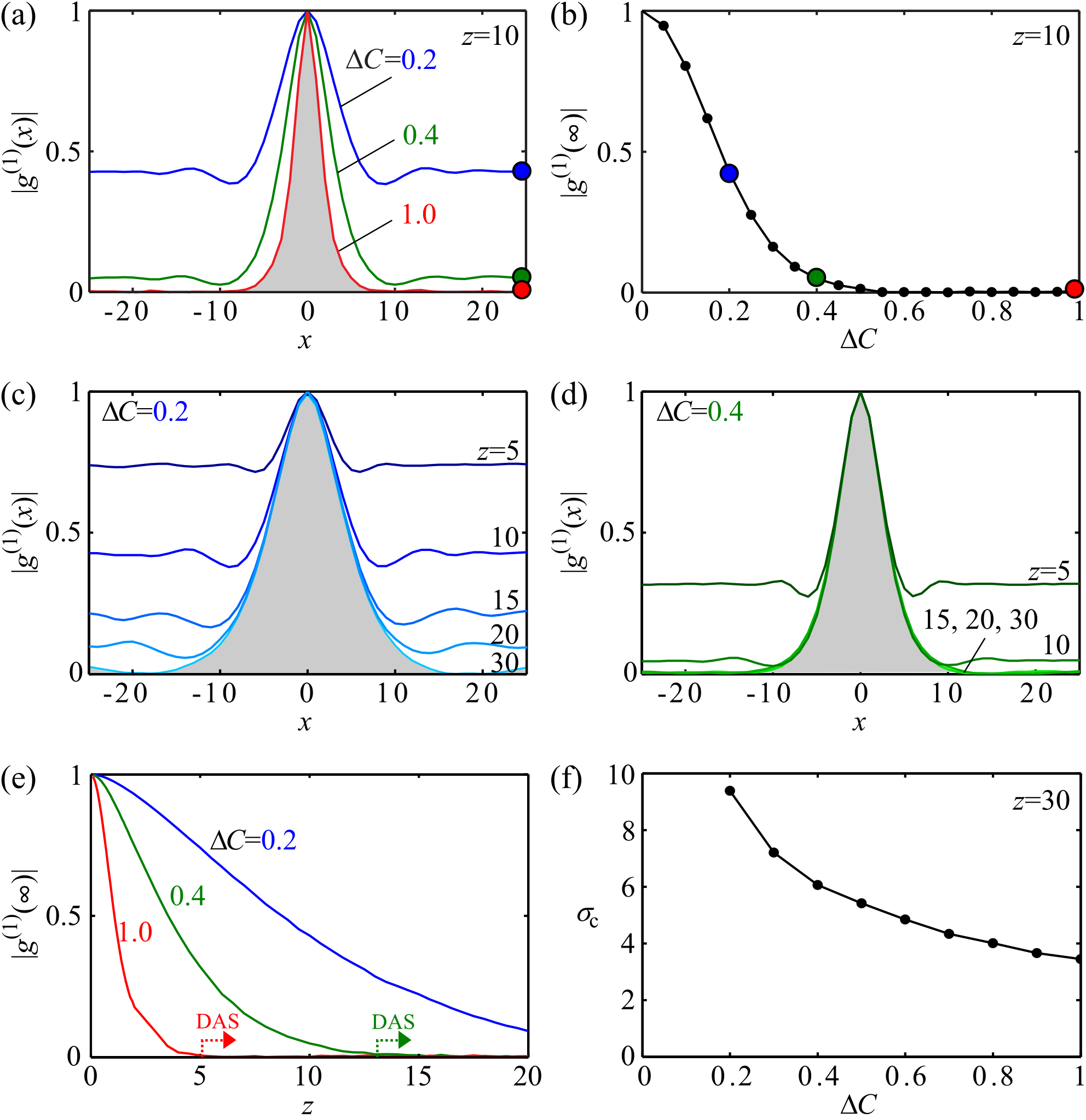} \vspace{-0.2cm}
	\caption{\label{Figure4}  \textbf{Transverse coherence for 1D discrete Anderson speckle.} \\ (a) Magnitude of $g^{(1)}(x)$ for 1D arrays for various disorder levels $\Delta C$ at propagation distance $z\!=\!10$. (b) The long-range-order coherence pedestal $|g^{(1)}(\infty)|$ as a function of $\Delta C$ at $z\!=\!10$. The circles in (b) correspond to the same values of $\Delta C$ in (a). (c,d) The magnitude of $g^{(1)}(x)$ at various $z$ for (c) $\Delta C\!=\!0.2$ and (d) $\Delta C\!=\!0.4$. The pedestal decreases with $z$ and $g^{(1)}(x)$ becomes stationary with respect to further propagation. (e) $|g^{(1)}(\infty)|$ as a function of $z$ at various $\Delta C$. (f) Transverse coherence width $\sigma_{\mathrm{c}}$ as a function of $\Delta C$ at $z\!=\!20$. All areas shaded in gray, and also the dashed arrows, indicate the onset of the discrete Anderson speckle (DAS) regime.}\vspace{-0.5cm}
\end{figure}

\begin{figure}[t!]
	\centering\includegraphics[scale=1]{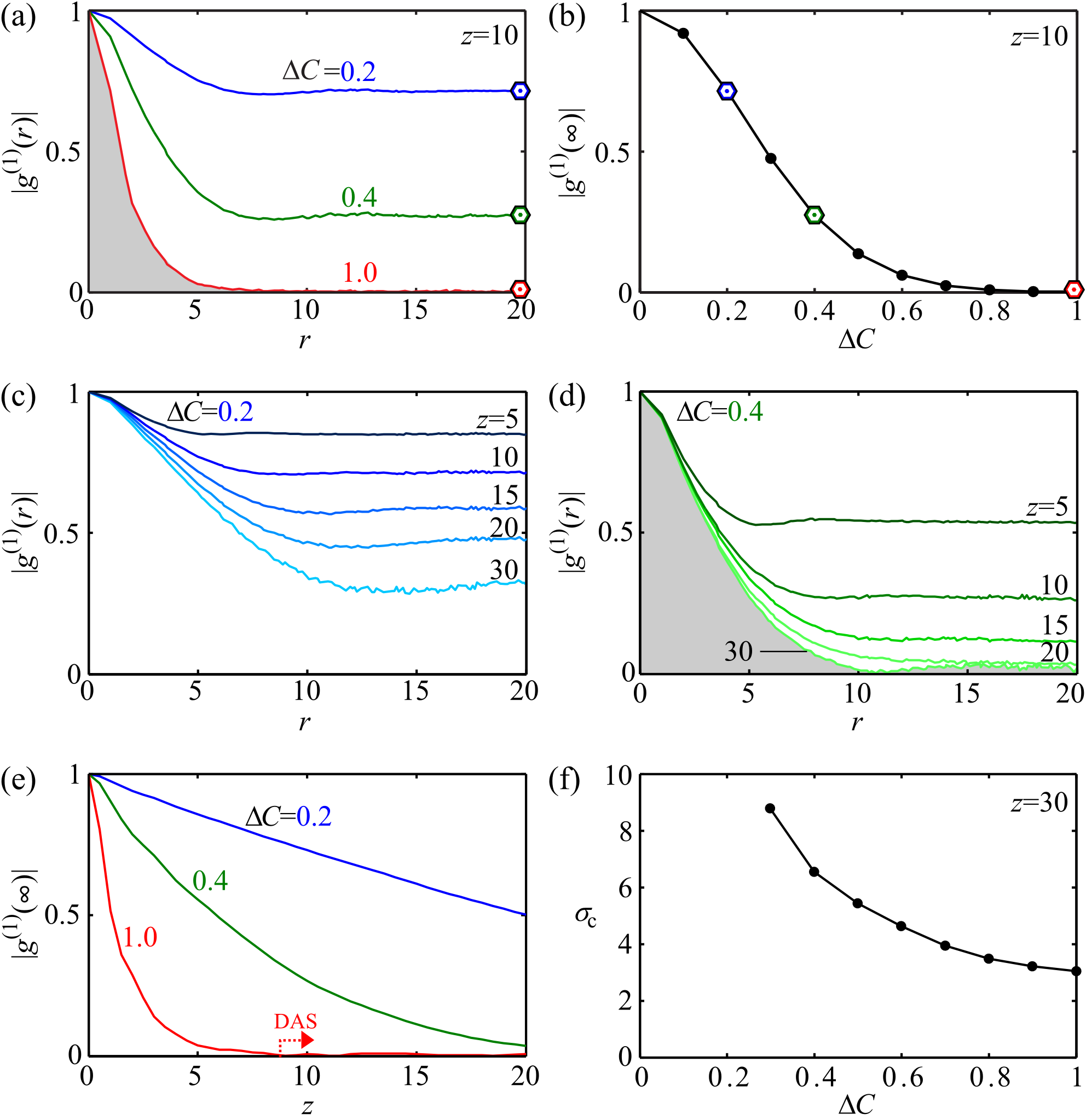} \vspace{-0.2cm}
	\caption{\label{Figure5}  \textbf{Transverse coherence for 2D discrete Anderson speckle.} \\ (a) Magnitude of $g^{(1)}(r)$ for 2D arrays for various disorder levels $\Delta C$ at propagation distance $z\!=\!10$. (b) The long-range-order coherence pedestal $|g^{(1)}(\infty)|$ as a function of $\Delta C$ at $z\!=\!10$. The hexagons in (b) correspond to the same values of $\Delta C$ in (a). (c,d) The magnitude of $g^{(1)}(r)$ at various $z$ for (c) $\Delta C\!=\!0.2$ and (d) $\Delta C\!=\!0.4$. The pedestal height decreases with $z$ and $g^{(1)}(x)$ becomes stationary with respect to further propagation. (e) $|g^{(1)}(\infty)|$ as a function of $z$ at various $\Delta C$. (f) Transverse coherence width $\sigma_{\mathrm{c}}$ as a function of $\Delta C$ at $z\!=\!20$. All areas shaded in gray, and also the dashed arrows, indicate the onset of the discrete Anderson speckle (DAS) regime.}\vspace{-0.5cm}
\end{figure}

We have carried out an extensive computational exploration of the coherence properties of light propagating in Anderson lattices. Figures~\ref{Figure4}(a) and~\ref{Figure5}(a) depict the magnitudes of $g^{(1)}(x)$ and $g^{(1)}(r)$ for 1D and 2D lattices, respectively, revealing a non-zero pedestal $|g^{(1)}(\infty)|$ riding on which is a finite-width distribution. This pedestal $|g^{(1)}(\infty)|$ signifies the survival of long-range transverse order; that is, some level of transverse correlation is maintained regardless of the separation between the pair of waveguides. Indeed, $|g^{(1)}(\infty)|$ decreases monotonically with $\Delta C$  until it vanishes altogether at a threshold $\Delta C$ value [Fig.~\ref{Figure4}(b) for 1D and Fig.~\ref{Figure5}(b) for 2D].

It is useful at this point to compare the coherence of discrete speckle described above to that of conventional speckle produced in the arrangement shown in Fig.~\ref{Figure1}(a). The random component of the screen phase $\phi$ is typically a Gaussian process with zero mean, variance $\sigma_{\phi}^2$, and spatially invariant transverse correlation of width $x_{\mathrm{c}}$, which we take as a transverse unit length in analogy to the unit separation between the waveguides on a lattice. During propagation along $z$, the field passes through two regimes. In the first regime where $z\!<\!2{N_{\mathrm{c}}x_{\mathrm{c}}^{2}}/{\lambda}$ ($N_{\mathrm{c}}$ is the size of the illuminating beam in units of $x_{\mathrm{c}}$ and $\lambda$ is the wavelength), the coherence properties do not change with $z$. Interestingly, the coherence function $g^{(1)}(x)$ for conventional speckle contains a pedestal associated with the specular component of the field when the thin phase screen has small $\sigma_{\phi}^{2}$ \cite{Goodman2007}, in analogy to the pedestal resulting from ballistic propagation in its discrete counterpart for small $\Delta C$ [Figs.~\ref{Figure4}(a) and ~\ref{Figure5}(a)]. In conventional speckle, the pedestal height drops gradually with increased $\sigma_{\phi}^{2}$ for fixed $z$ [similarly to the behavior of $|g^{(1)}(\infty)|$ with $\Delta C$ in Figs.~\ref{Figure4}(b) and ~\ref{Figure5}(b)], and gradually vanishes as the field leaves this regime, i.e., $z\!>\!2{N_{\mathrm{c}}x_{\mathrm{c}}^{2}}/{\lambda}$. In the far field, $g^{(1)}(x)$ becomes the Fourier transform of the illumination spot and the grain size increases continuously with $z$ in accordance with the van Cittert-Zernike theorem [Fig.~\ref{Figure1}(c)].

A distinction between `near-' and `far-field' may be similarly made for discrete speckle based on the disappearance of the pedestal $g^{(1)}(\infty)$. For small distances, $g^{(1)}(\infty)$ is non-zero and the discrete speckle undergoes dynamical changes upon propagation as shown in Fig.~\ref{Figure4}(c)-(d). However, for a given disorder level $\Delta C$, the pedestal vanishes after some distance $z\!>\!\frac{5}{\Delta C}$ [Fig.~\ref{Figure4}(e)] that we determined empirically -- which we take as an indication that the `far-field' has been reached ($z\!>\!\frac{10}{\Delta\beta}$ for arrays with diagonal disorder). Beyond this axial distance, $g^{(1)}(x)$ is stationary and the grain size freezes. This observation is a glaring departure from conventional speckle where grain size increases upon propagation in the far field. We call discrete speckle in this regime \textit{discrete Anderson speckle}, since we will show later that the transverse coherence width $\sigma_{\mathrm{c}}$ here is dictated by the localization length $\sigma_{\mathrm{s}}$. We define the coherence width $\sigma_{\mathrm{c}}$ (or grain size) as the full-width at half-maximum (FWHM) of the steady-state $|g^{(1)}(x)|$ (that is, in the far-field where the pedestal $g^{(1)}(\infty)$ disappears). We find that $\sigma_{\mathrm{c}}$ decreases monotonically with $\Delta C$ as shown in Fig.~\ref{Figure4}(f). In the near-field, the pedestal in effect reduces the speckle grain size by screening the steady-state $|g^{(1)}(x)|$. Freezing of the coherence function with propagation takes place slower in 2D [Fig.~\ref{Figure5}].

\begin{figure}[t!]
	\centering\includegraphics[scale=1.1]{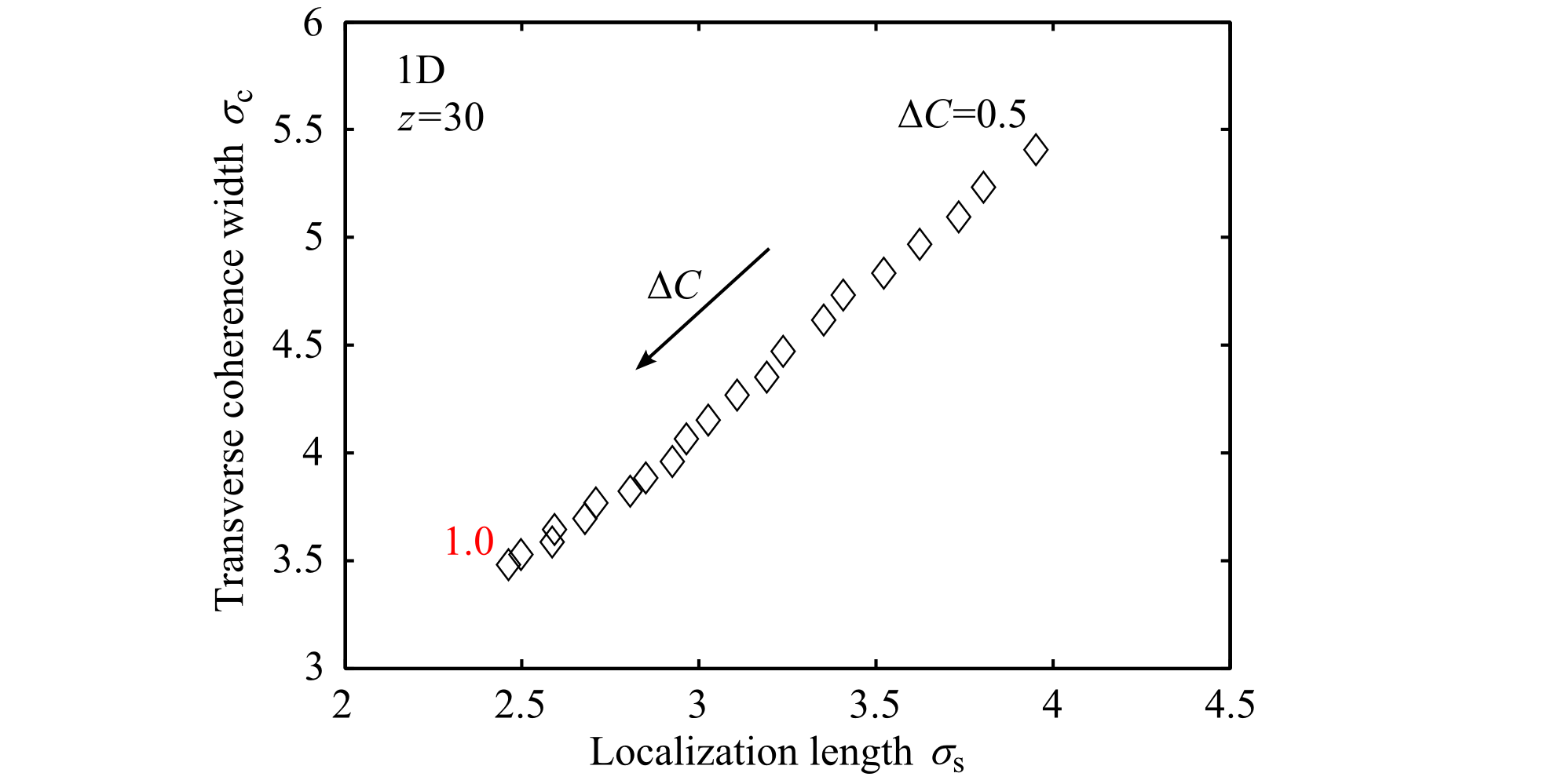} \vspace{-0.2cm}
	\caption{\label{Figure6}  \textbf{Correlation between transverse coherence and localization in the discrete Anderson speckle regime.} \\ Correlation between $\sigma_{\mathrm{s}}$ and $\sigma_{\mathrm{c}}$ with varying disorder level at $z\!=\!20$. Here $\sigma_{\mathrm{s}}$ is the FWHM of the mean PSF and $\sigma_{\mathrm{c}}$ is the FWHM of the degree of transverse coherence $|g^{(1)}(x)|$. The axial distance $z\!=\!20$ is selected such that the discrete Anderson speckle regime (where $|g^{(1)}\!(\infty)|\!\approx\!0$) has been reached for all disorder levels from $\Delta C\!=\!0.5$ to 1.}\vspace{-0.5cm}
\end{figure}

To uncover the physics underlying this disorder-induced freezing of the grain size in the discrete Anderson regime, we compare $\sigma_{\mathrm{c}}$ for \textit{uniform} illumination to the localization length $\sigma_{\mathrm{s}}$ resulting from a \textit{single}-waveguide excitation. For this comparison, we re-define $\sigma_{\mathrm{s}}$ as the FWHM of the mean PSF. We find that these two very different quantities are in fact linearly proportional $\sigma_{\mathrm{c}}\!\approx1.3\sigma_{\mathrm{s}}$ [Fig.~\ref{Figure6}]. This may be understood by noting that in the presence of disorder the PSF is a random function with finite average width \cite{Martin2011a, Abouraddy2012b, DiGiuseppe2013}. Light emerging from waveguides separated by a distance greater than the PSF width are likely to have passed through non-overlapping paths of the random array, and should therefore be uncorrelated. We will present below a general analytical argument that establishes the relationship between $\sigma_{\mathrm{c}}$ for extended illumination to $\sigma_{\mathrm{s}}$ for a single-waveguide excitation.

\section{Discrete Anderson Localization:  Axial Coherence}	
\begin{figure}[t!]
	\centering \includegraphics[scale=1]{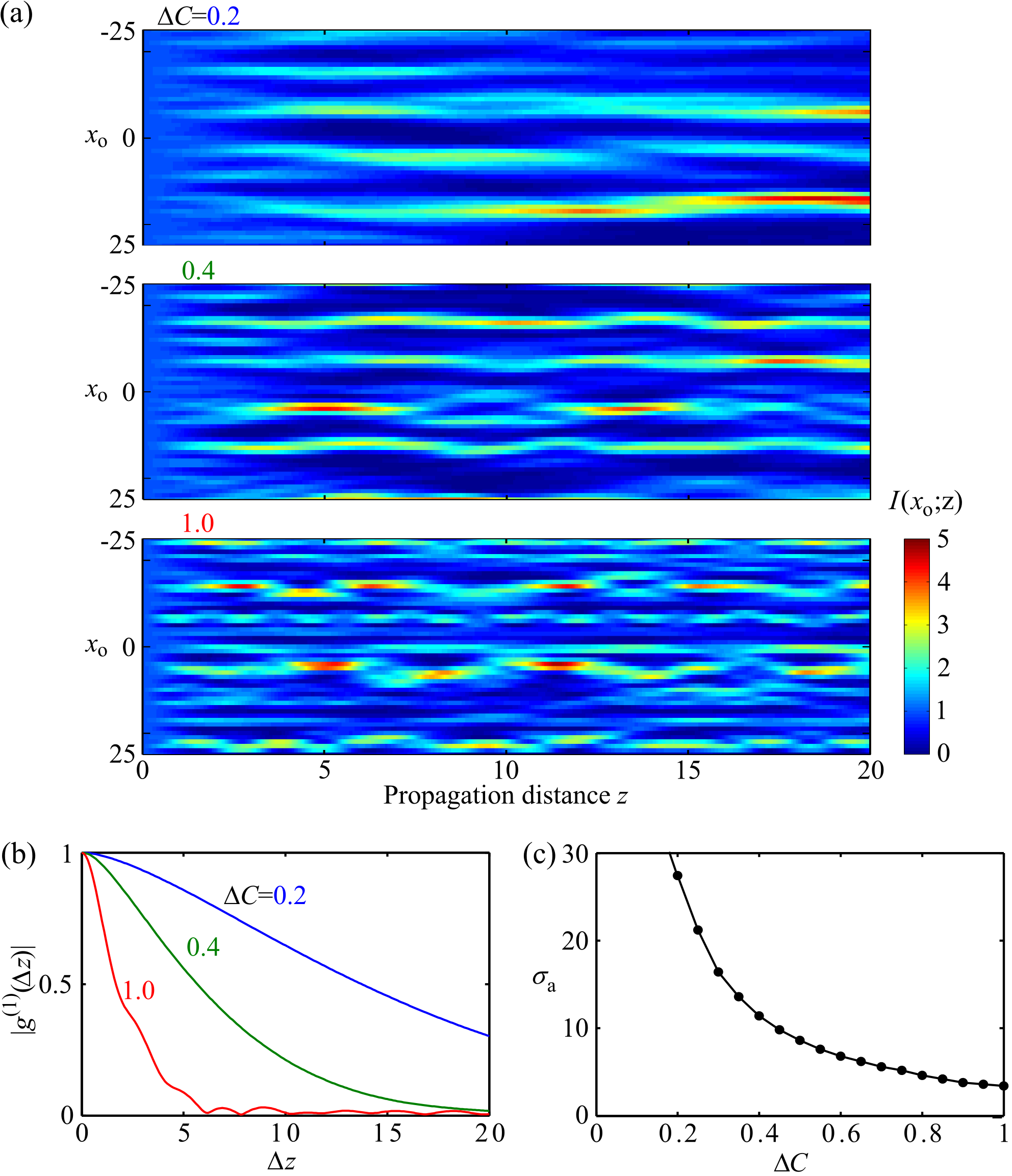} \vspace{-0.2cm}
	\caption{\label{Figure7}  \textbf{Axial coherence in 1D discrete Anderson speckle.} \\ (a) Axial evolution of the intensity in individual realizations of 1D lattices with different $\Delta C$. (b) The amplitude of the axial coherence function $|g^{(1)}(\Delta z)|$ and (c) the axial coherence depth $\sigma_{\mathrm{a}}$ for different $\Delta C$.}\vspace{-0.5cm}
\end{figure}

Further insight may be drawn from a detailed examination of the axial coherence propagation dynamics. We plot $I(x_{\mathrm{o}};z)\!=\!|E(x_{\mathrm{o}};z)|^{2}$ for three realizations at $\Delta C\!=\!0.2,0.4,1.0$ in Fig.~\ref{Figure7}(a). The longitudinal freezing of the transverse discrete speckle is evident for all three cases in the far field, resulting in axial filamentation of the intensity distribution -- corresponding to the non-overlapping uncorrelated paths along the disordered lattice mentioned above. Evaluation of the axial coherence function $G^{(1)}(z,\Delta z)\!=\!\Sigma_{x_{\mathrm{o}}}\langle E^{*}(x_{\mathrm{o}};z)E(x_{\mathrm{o}};z+\Delta z)\rangle$ reveals that it is in fact \textit{independent} of $z$ altogether. The normalized axial degree of coherence $|g^{(1)}(\Delta z)|$ decays with $\Delta z$ at a rate proportional to the disorder level [Fig.~\ref{Figure7}(b)], so that its FWHM or \textit{axial} coherence depth $\sigma_{\mathrm{a}}$ drops with disorder [Fig.~\ref{Figure7}(c)]. This behavior is stationary along $z$.  Finally, a unique aspect of the features described in this Section is that they are evident in \textit{individual} realizations, unlike observations of Anderson localization that necessitate \textit{ensemble averaging}.

We have found that the transverse coherence width $\sigma_{\mathrm{c}}$ reaches a steady state in the discrete Anderson speckle regime and the statistical homogeneity renders it independent of transverse position $x$. Furthermore, the axial coherence depth $\sigma_{\mathrm{a}}$ for a fixed disorder level is independent of axial position $z$ (and is primarily due to dephasing; see Figs. S3-S5 in the Supplement). By combining these findings concerning transverse and axial coherence in disordered lattices, we conclude that a coherence `voxel' of fixed volume exists everywhere along the lattice in the discrete Anderson speckle regime. The volume of this coherence voxel depends solely on the disorder level $\Delta C$. This behavior is once again a dramatic departure from that of conventional speckle where the transverse coherence growth in the far field is dictated by the van Citter-Zernike theorem, and this growth in transverse coherence length is accompanied by a reduction in the axial coherence length.

\section{Analytical Model}

We have shown numerically that the fingerprint of localization exists in the fluctuations of the discrete speckle emerging from Anderson lattices for an extended coherent input. It may be initially surprising that a link exists between the \textit{localization length} (typically associated with a point excitation and averaging over output intensity) and the \textit{transverse coherence width} (associated with an extended input and averaging over field products for pairs of waveguides); see Fig.~\ref{Figure6}. Our goal here is to link the \textit{extended-illumination} scheme that has been our focus [Fig.~\ref{Figure1}(b)] with the more usual \textit{single-waveguide} excitation strategy [Fig.~\ref{Figure2}(a,b)]. To elucidate this link, we adapt to our setting a conceptual scheme from quantum optics known as `Klyshko's advanced-wave picture' \cite{Strekalov1995, Klyshko1994}, which is also of use to classical fields. This scheme allows for the identification of correlation functions of an \textit{extended field} traversing an optical system with the field or intensity of a double-pass configuration (backward then forward) of a \textit{point source} through the same system.

\begin{figure*}[t]
	\centering \includegraphics[scale=1.1]{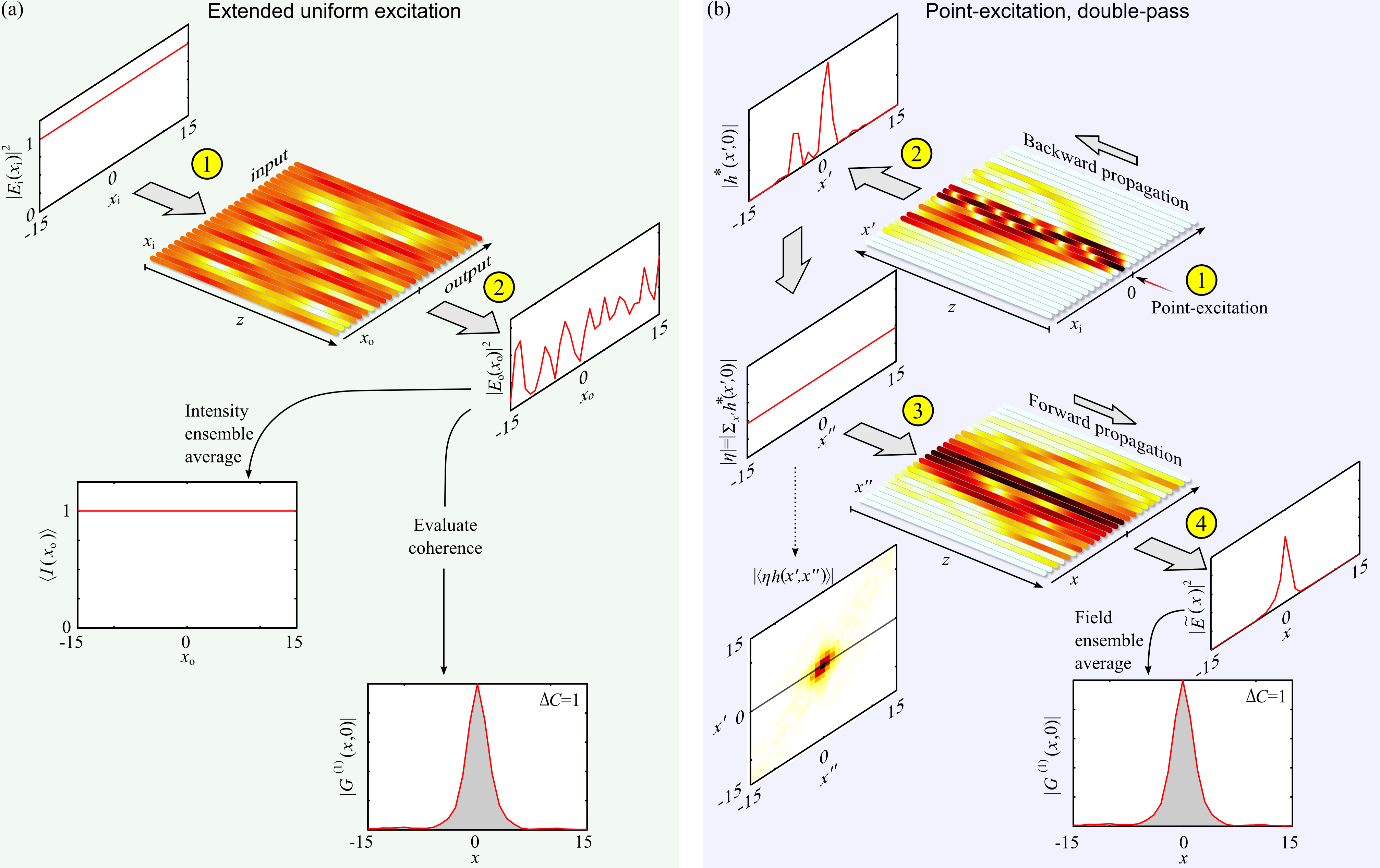} \vspace{-0.2cm}
	\caption{\label{Figure8}  \textbf{Heuristic model linking the transverse coherence width to the localization length in an Anderson lattice.} \\ (a) Schematic for  extended uniform excitation $E_{\mathrm{i}}(x_{\mathrm{i}})\!=\!1$ in an array of waveguides resulting in an output field  $E_{\mathrm{o}}(x_{\mathrm{o}})$   having  a narrow transverse coherence function with no pedestal. Here $\Delta C\!=\!1$ and $z$ is taken such that we are in the discrete Anderson speckle regime.  Ensemble averaging of the output intensity  $\langle|I(x_{\mathrm{o}})|^{2}\rangle$  yields a constant. (b) A representation of Klyshko's advanced-wave picture in which an unfolded cascade of systems is excited at a single point ($x_{\mathrm{i}}\!=\!0$) and whose output may be put in correspondence with that of the extended illumination configuration in (a).  The field propagates backward through the disordered lattice (as a result of the conjugation in Eq.~\ref{eq:G1_extendedfield}) and a spatial average of the output field $\eta\!=\!\sum_{x'}\!h^{*}(0,x')$ is evaluated. An extended uniform field with random complex amplitude $\eta$ propagates forward through the same realization of the lattice to produce an output field $\tilde{E}(x)$ whose ensemble average $\langle\tilde{E}(x)\rangle$ corresponds to the coherence function in (a). We also plot the function $\langle\eta h(x,x'')\rangle$ for reference. $\Delta C\!=\!1$ and ensemble size is $10^4$ in (a)-(b).} \vspace{-0.5cm}
\end{figure*}

We start by depicting in Fig.~\ref{Figure8}(a) the 1D scenario we have investigated in this paper, whereupon an extended coherent field traverses a random lattice ($\Delta C\!=\!1$). Averaging the output intensity $|E_{\mathrm{o}}(x_{\mathrm{o}})|^{2}$ over multiple realizations yields a constant distribution with no localization signature [Fig.~\ref{Figure2}(c)]. Nevertheless, computing the spatially stationary coherence function $G^{(1)}(0,x)$ by averaging over products of fields from pairs of waveguides separated by $x$ yields a localized function (independently of $x_{\mathrm{o}}$) of width $\sigma_{\mathrm{c}}$. Referring to Eq.~\ref{CoherenceFunctionDefinition}, we write $G^{(1)}(0,x)$ as
\begin{equation}\label{eq:G1_extendedfield}
G^{(1)}(0,x)=\langle\,\overbrace{\sum_{x''} h(x,x'')}^{\mathrm{forward}}\!\!\!\!\!\!\underbrace{\sum_{x'}}_{\mathrm{averaging}}\!\!\!\!\!\!\overbrace{h^{*}(0,x')}^{\mathrm{backward}}\,\rangle.
\end{equation}
This equation can be interpreted in light of the Klyshko advanced-wave picture as a \textit{cascade} of the three steps illustrated in Fig.~\ref{Figure8}(b). First, a point excitation at $x_{\mathrm{i}}\!=\!0$ propagates backward through the system $h$ to the $x'$ plane, as dictated by the conjugation operation. Second, the output field from this backward propagation is spatially averaged over $x'$ to yield the complex random variable $\eta\!=\!\sum_{x'}h^{*}(0,x')$, which is then equally distributed over points $x''$ in the input plane for a second pass forward through the \textit{same realization} of the system $h$. Third, the uniform extended field of amplitude $\eta$ propagates forward through $h$ to produce an output random field $\tilde{E}(x)\!=\!\eta\sum_{x''}h(x,x'')$. Ensemble averaging results in $\langle\tilde{E}(x)\rangle\!=\!G^{(1)}(0,x)$ per Eq.~\ref{CoherenceFunctionDefinition} and Eq.~\ref{eq:G1_extendedfield}.

Let us examine the third step in this cascade, the forward pass. Each waveguide at position $x''$ is fed with a noisy field having complex random amplitude $\eta$ with zero mean. The ensemble average of the output field in the $x$-plane contributed by each waveguide is $\langle\eta h(x,x''\!)\!\rangle$. While the ensemble average of $\langle\eta\rangle$ and $\langle h(x,x''\!)\!\rangle$ (for high disorder levels) is \textit{each} zero, the average of their product need not be so since \textit{both random variables are generated by the same realization of the disordered lattice}. Indeed, since $\eta$ is generated by the random lattice environment in the vicinity of $x\!=\!0$ in the Anderson localization limit, then it correlates only with $h$ in the same vicinity, while remaining uncorrelated $\langle\eta h(x,x''\!)\!\rangle\!\sim\!0$ when $h$ is evaluated away from the origin, as shown in Fig.~\ref{Figure8}(b). Consequently, only a few waveguides in the vicinity of $x''\!\!=\!0$ contribute to the forward pass. Since $h$ produces a localized output for a point excitation, the few-waveguide excitation here results in a slightly broader localized spot whose width is $\sigma_{\mathrm{c}}$ (resulting from the convolution of the impulse response function with the width of the distribution in Fig.~\ref{Figure8}(b)). We have thus established on these grounds that $\sigma_{\mathrm{c}}$ is intimately linked with the localization length $\sigma_{\mathrm{s}}$, but is expected to be slightly larger -- as was shown numerically in Fig.~\ref{Figure6}.

We next proceed to an analytical model of discrete Anderson speckle based on modal analysis. Using the eigenmodes of the lattice coupling matrix, we justify (1) the freezing of the transverse coherence width $\sigma_{\mathrm{c}}$ (and hence the speckle grain size) once the discrete Anderson speckle regime is reached, and (2) the independence of the axial coherence depth $\sigma_{\mathrm{a}}$ from axial position $z$.

\subsection{Origin of the freezing of the transverse coherence width}
We analyze the propagation of the field along an Anderson lattice in terms of the eigenmodes and eigenvalues of the Hermitian coupling matrix $\hat{\mathbf{H}}$ that is defined by the equation of dynamics in Eq.~\ref{eq:general} by writing
\begin{equation}
i\frac{d\mathbf{E}(z)}{dz}+\hat{\mathbf{H}}\mathbf{E}=0,
\end{equation}
where $\mathbf{E}$ is a vector of length $2N+1$ containing the field amplitudes in the waveguides, and $\hat{\mathbf{H}}$ is a real symmetric (and hence Hermitian) matrix with the wave numbers along the diagonal and coupling coefficients off the diagonal. If the eigenmodes and eigenvalues of $\hat{\mathbf{H}}$ are $\phi_n(x)$ and $b_n$, respectively, then since $h\!=\!e^{i\hat{\mathbf{H}}z}$, the eigenvalue problem is defined for the impulse response function as
\begin{equation}
\sum_{x'}h(x,x';z)\phi_{n}(x')=e^{ib_{n}z}\phi_{n}(x),
\end{equation}
such that the impulse response function may be expressed as
\begin{equation}\label{PSFEigenmodes}
h(x_\mathrm{o},x_\mathrm{i};z)=\sum_n e^{ib_nz} \phi_n(x_\mathrm{o}) \phi_n(x_\mathrm{i}).
\end{equation}
We have made use of the fact that the eigenmodes are real since $\hat{\mathbf{H}}$ is real and symmetric. Using this definition, we recast the joint transverse-axial coherence function in terms of $\phi_n(x)$ and $b_n$,
\begin{eqnarray}\label{eq:g1mostgeneral}
&G^{(1)}&(0,x;z,z\!+\!\Delta z)=\sum_{x',x''}\langle h^{*}(0,x';z)h(x,x'';z+\Delta z)\rangle\nonumber\\&=&\sum_{x',x''}\sum_{n,m} \langle \phi_n(0)\phi_n(x')\phi_m(x)\phi_m(x'') e^{i(b_n-b_m)z}e^{-ib_m\Delta z} \rangle.
\end{eqnarray}

The freezing of the speckle grain size in the discrete Anderson speckle regime is realized at large propagation distances $z$ when the following condition is satisfied: $\mathrm{Std}\{b_n\}z\!\gtrsim\!2\pi$; here $\mathrm{Std}\{\cdot\}$ is the standard deviation. We expect that $\mathrm{Std}\{b_n\}$ is proportional to $\Delta C$, such that the distance $z$ that satisfies this condition is inversely proportional to $\Delta C$. In the case of off-diagonal disorder, which we have considered here, the eigenvalue $b_{0}$ is excluded from this condition since it remains deterministic with value 0 \cite{Evangelou2003}. This exclusion is not required in the case of diagonal disorder which is described in the Supplement. We have found numerically that this limit in lattices with off-diagonal disorder is attained when $z \Delta C \!>\! 5$, which we have taken to define the discrete Anderson speckle regime.

When the condition $\mathrm{Std}\{b_n\}z\!\gtrsim\!2\pi$ is met, the difference $b_n\!-\!b_m$ when $n\neq m$ has same order of magnitude as this standard deviation, but is equal to zero when $n\!=\!m$, therefore implying that upon ensemble averaging, the impact of the exponential term in Eq.~\ref{eq:g1mostgeneral} is $e^{i(b_n-b_m)z}\rightarrow\delta_{n,m}$. Thus, setting $\Delta z\!=\!0$ in the axial regime where $z \Delta C \!>\! 5$, Eq.~\ref{eq:g1mostgeneral} reduces to
\begin{equation}
G^{(1)}(0,x;z,z)\!=\!\sum_{x',x''}\sum_{n} \langle \phi_n(0)\phi_n(x)\phi_n(x')\phi_n(x'') \rangle.
\end{equation}
This equation implies that in the discrete Anderson speckle regime the transverse coherence is a function of the separation $x$ but not the axial distance $z$ -- as demonstrated numerically in Fig.~\ref{Figure4}.

\subsection{Independence of axial coherence depth from axial position}
In considering the axial coherence along the lattice, we make use of the transverse stationarity of the lattice and consider a single lattice site $x$ in Eq.~\ref{eq:g1mostgeneral}, whereupon the axial coherence function is
\begin{eqnarray}
G^{(1)}\!\!\!\!\!\!\!\!\!&&(x,x;z,z+\Delta z)\nonumber\\
&=&\sum_{x',x''}\sum_{n,m}\langle\phi_n(x)\phi_n(x')\phi_m(x)\phi_m(x'')e^{i(b_n-b_m)z}e^{-ib_m\Delta z}\rangle.
\end{eqnarray}
By taking a spatial average over $x$, we obtain a simplified relation
\begin{equation}
\sum_{x}G^{(1)}(x,x;z,z+\Delta z)=\sum_{x',x''}\sum_{n}\langle\phi_n(x')\phi_n(x'')e^{-ib_n\Delta z}\rangle,
\end{equation}
in which we used $\sum_{x}\!\phi_n(x)\phi_m(x)\!\!=\!\!\delta_{n,m}$. Consequently, the axial coherence function averaged over the transverse coordinate is altogether independent from $z$. However, since $G^{(1)}\!(x,x;z,z+\Delta z)$ is stationary in $x$, its statistical properties are the same as those of $\sum_{x}\!G^{(1)}\!(x,x;z,z+\Delta z)$. Therefore, the axial coherence function is independent of $z$, and as a result its width $\sigma_{\mathrm{a}}$ is also independent of $z$ and relies only on $\Delta z$ -- as demonstrated numerically in Fig.~\ref{Figure7}.

\section{Conclusion}
We have investigated the evolution of a set of mutually coherent waves traveling through 1D and 2D disordered lattices of coupled waveguides. The emerging wave forms discrete speckle that is statistically homogeneous with random intensity distribution on the lattice sites. The disordered lattice structure that results in Anderson localization when a single waveguide is excited exhibits in the case of an extended excitation a complete freezing of the discrete speckle grain size after reaching a steady-state, unlike the usual growth observed in \textit{conventional} speckle -- a regime we refer to as discrete \textit{Anderson} speckle. Moreover, axial and transverse coherence are independent of position, resulting in a coherence voxel of fixed volume independent of its transverse and axial position on the lattice. These results are applicable to a broad host of photonic systems in which disorder may impact coupling between discrete elements \cite{Schwartz2007a, Lahini2008a, Martin2011a, Topolancik2007, Karbasi2014, Mookherjea2008, Abouraddy2012b, DiGiuseppe2013, Crespi2013}. While we have studied second-order field correlations on a discrete lattice, the new behavior reported here signposts important vistas to be investigated in the context of higher-corder correlations and photon statistics \cite{Lahini2011a}.

Finally, the correspondence between the propagation of light and that of a quantum particle on discrete lattices \cite{Christodoulides2003a} has led recently to fruitful exchanges between optical and condensed matter physics  \cite{Morandotti1999a, Lederer2008, Iwanow2005, Peruzzo2010c, Shandarova2009a,  Rechtsman2013, Rechtsman2013b}. Our result, therefore, points to new regimes that may be investigated in other physical systems, ranging from Bose-Einstein condensates \cite{Billy2008a} to acoustic lattices \cite{Hu2008}, where Anderson localization takes place owing to interference of random waves. 

\section*{Acknowledgments}
The authors thank the Stokes Advanced Research Computing Center at the University of Central Florida for access to the high-performance computing cluster.

\vspace{0.3cm}
\noindent
See Supplement for supporting content.

%===============REFERENCES==============

\newpage 

\begin{widetext}
\section*{supplementary material} 
In this supplementary material,  we provide additional computational results. Specifically, we carry out a computational study to support the key results reported in the main text, but applied to 1D and 2D waveguide arrays with \textit{diagonal} disorder. After briefly reviewing the difference between diagonal and off-diagonal disorder in our calculations, we proceed in the same order followed in the main text. In addition, we clarify the source for the dephasing of the optical field with propagation in disordered arrays for both disorder classes. The computations indicate that there are no qualitative differences between the behavior exhibited by lattices with  diagonal or off-diagonal disorder.
\end{widetext}

\section*{Model}
Consider an array of parallel waveguides on a 1D lattice with evanescent nearest-neighbor-only coupling. The field propagation is described by the coupled equations
\begin{equation}\label{eq:general}
i\frac{dE_{x}(z)}{dz}+\beta_{x} E_{x}+C_{x,x-1} E_{x-1}+C_{x,x+1} E_{x+1}=0,
\end{equation}
where $E_{x}(z)$ is the complex optical field in the ${x}^{\text{th}}$ waveguide $({x}\!=\!-N,\ldots,N)$ at axial position $z$,  $\beta_{x}$ is the propagation constant of waveguide $x$, and $C_{x,x+1}$ is the coupling coefficient between adjacent waveguides $x$ and $x+1$. These equations may be written as $$i\frac{d\mathbf{E}}{dz}+\hat{\mathbf{H}}\mathbf{E}=0,$$ where $\mathbf{E}(z)$ is a vector with components $\{E_{x}(z)\}_{x=-N}^{N}$ and $\hat{\mathbf{H}}$ is the Hermitial coupling matrix given by
\begin{equation}\label{eq:matrix2}
\hat{\mathbf{H}}=
\begin{pmatrix}
\ddots	&\ddots		&0		&\dots		&0 \\
\ddots	&\beta_{x-1}	&C_{x-1,x}	&\ddots		&\vdots \\
0		&C_{x,x-1}		&\beta_{x}	&C_{x,x+1}	&0\\
\vdots	&\ddots		&C_{x+1,x}	&\beta_{x+1}	&\ddots\\
0		&\dots		&0		&\ddots		&\ddots
\end{pmatrix}
\end{equation}
The evolution of the input field $E_\mathrm{i}(x_\mathrm{i})\!=\!E_{x_\mathrm{i}}(z\!=\!0)$ to the output $E_\mathrm{o}(x_\mathrm{o})\!=\!E_{x_{o}}(z)$ may be written as
\begin{equation}\label{eq:inoout}
E_\mathrm{o}(x_\mathrm{o})=\sum_{x_\mathrm{i}} h(x_\mathrm{o},x_\mathrm{i})E_\mathrm{i}(x_\mathrm{i}),
\end{equation}
where $h(x_\mathrm{o},x_\mathrm{i})$ corresponds to the matrix elements of $e^{i\hat{\mathbf{H}}z}$ and represents the impulse response function of the system. The point spread function (PSF) is $|h(x_\mathrm{o},x_\mathrm{i})|^2$ and corresponds to the output intensity. This formulation is readily extended  to 2D lattices.

In this supplement, we focus on \textit{diagonal} disorder, which is characterized by constant coupling coefficients ($C_{x,x+1}=\overline{C}$) and random propagation constants $\beta_x$ having a uniform probability distribution of mean $\overline{\beta}$ and half width $\Delta \beta$. We use the same normalization scheme used in the main text. The coupling coefficients are written in units of their average $\overline{C}$, and the distance $z$ in units of the coupling length $\ell\!=\!1/\overline{C}$. Throughout, ${\Delta \beta}$ ranges from 0 to 2, and the lattice size is $N_{\mathrm{t}}\!=\!151$ for 1D and $N_{\mathrm{t}}\!\times\!N_{\mathrm{t}}\!=\!101\!\times\!101$ for 2D (hexagonal honeycomb) lattices  ($N_{\mathrm{t}}\!=\!2N+1$). The sizes are chosen such that reflections from the array boundaries do not affect the conclusions. Statistical averages are taken for an ensemble with $10^6$ and $10^4$ realizations for 1D and 2D simulations, respectively. Unless otherwise stated these parameters are  identical to those used for the calculations in the  off-diagonal disorder case presented in the main text unless otherwise stated.

\section*{Results}
\subsection*{Anderson Localization}
\begin{figure}[h!]
\includegraphics[width=8.4cm]{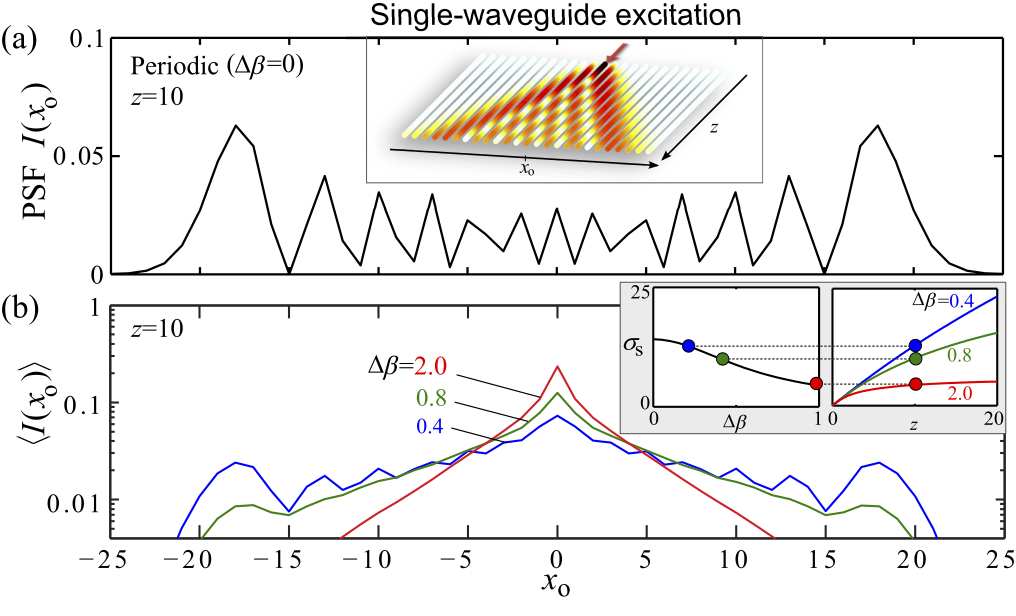}
\caption{\label{fig1a} (a) PSF $I(x_\mathrm{o})\!=\!|h(x_\mathrm{o},0)|^2$ at $z\!=\!10$  for a 1D periodic array with single-waveguide excitation at $x_\mathrm{i}\!=\!0$. Inset is a schematic of the configuration. (b)  Mean PSF $\langle I(x_\mathrm{o})\rangle\!=\!\langle|h(x_\mathrm{o},0)|^2\rangle$ for disordered 1D arrays. Insets show the localization length $\sigma_{\mathrm{s}}$ as a function of $\Delta\beta$ and $z$. For the values of $\sigma_{\mathrm{s}}$ in the insets, 21 points for $\Delta \beta$ and 200 for $z$ are chosen.}
\end{figure}

\begin{figure}
\includegraphics[width=8.4cm]{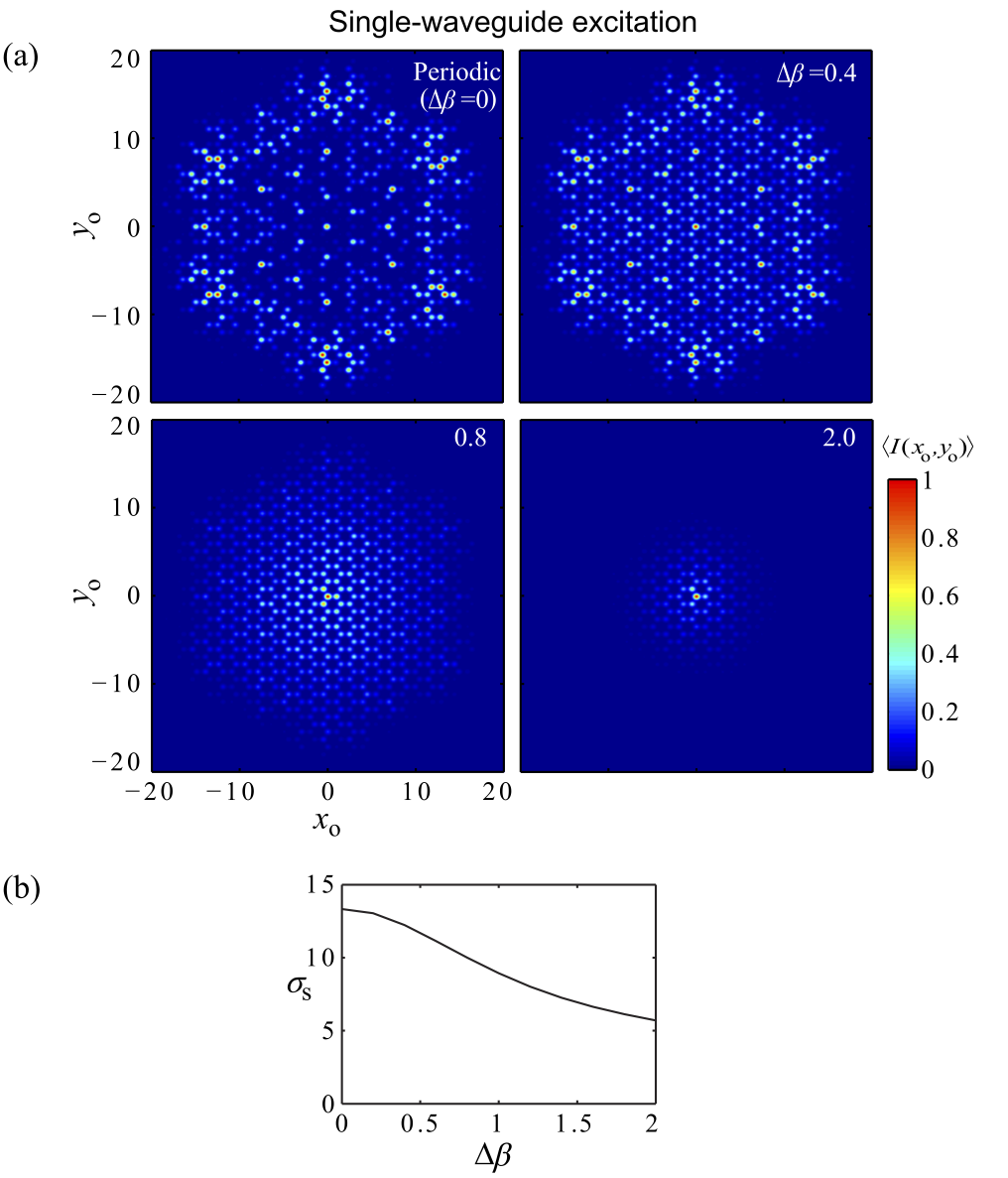}
\caption{\label{fig2a} (a) Mean PSF $\langle I(x_\mathrm{o},y_\mathrm{o})\rangle \!=\! \langle | h(x_\mathrm{o},0;y_\mathrm{o},0)|^2 \rangle$ for 2D hexagonal (honeycomb) arrays with increasing disorder (from left to right) at $z\!=\!10$. For clarity, each panel is normalized separately and convolved with a gaussian function of width $1.6$ for  visualization. (b) Localization radius $\sigma_{\mathrm{s}}$ as a function of disorder level $\Delta\beta$. For the values of $\sigma_{\mathrm{s}}$ shown, 11 points for $\Delta \beta$ are chosen.}
\end{figure}

We first report the spatial distribution of the mean output intensity for single-waveguide excitation in 1D and 2D waveguide lattices for selected disorder levels in Fig.~\ref{fig1a} and Fig.~\ref{fig2a}(a), respectively. In both cases, ballistic spread in periodic arrays leads to extended output states and gradually introducing disorder results in Anderson localization. The spatial width $\sigma_\mathrm{s}$ of the mean output intensity as a function of disorder level is depicted for 1D and 2D arrays in the inset of Fig.~\ref{fig1a}(b) and Fig.~\ref{fig2a}(b), respectively. 

\subsection*{Discrete Anderson Speckle - Transverse Coherence}	
\begin{figure}
\includegraphics[width=8.4cm]{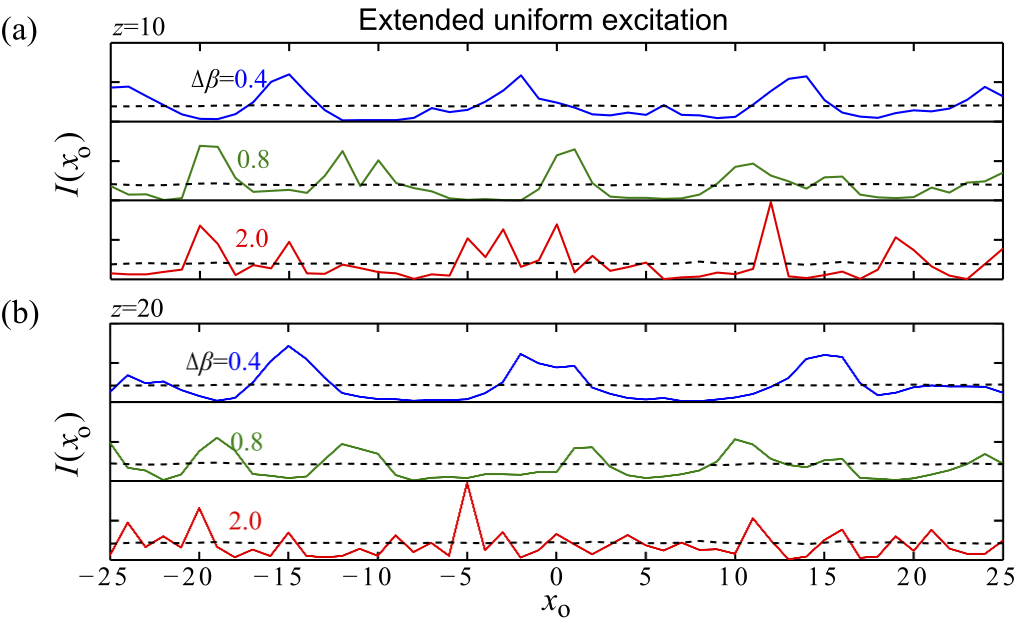}
\caption{\label{fig1b} (a) Realizations of the intensity distribution of discrete Anderson speckle at various disorder levels ($z\!=\!10$) for extended uniform coherent input light. The speckle grain size decreases with disorder. The dotted lines are ensemble averages that confirm the statistical homogeneity. (b) Same as (a) except that $z\!=\!20$.}
\end{figure}

\begin{figure}
\includegraphics[width=8.4cm]{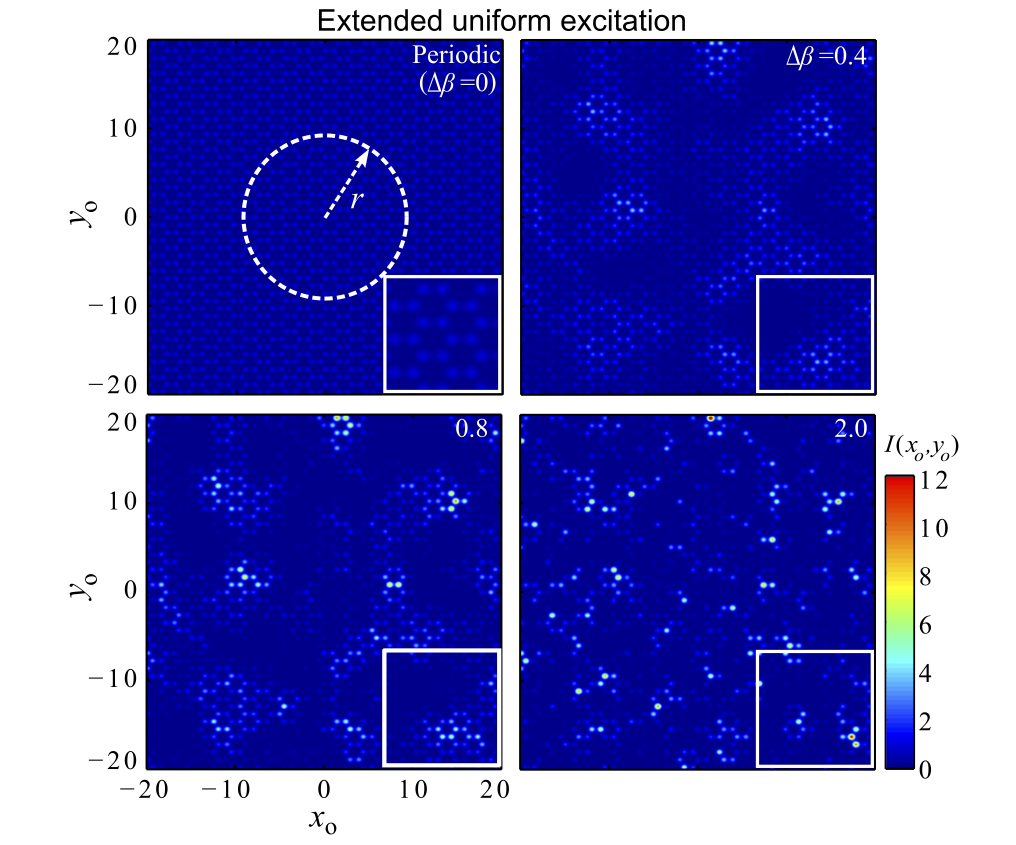}
\caption{\label{fig2b} Individual realizations of discrete Anderson speckle at various disorder levels corresponding to the panels in Fig.~\ref{fig2a}(a). The excitation is extended uniform coherent light. Note that the speckle grain size decreases with disorder. Insets are magnified by a factor of 2. Speckle contrasts are 0, 0.67, 1.17, 1.53 with increasing disorder level.}
\end{figure}

The focus in our paper is on the case of coherent uniform illumination extending across the whole array -- in \textit{diagonally} disordered Anderson lattices here. Examples of individual realizations for 1D and 2D lattices are shown in Fig.~\ref{fig1b}  and Fig.~\ref{fig2b}, respectively. Similar behavior in these discrete-speckle realizations is observed in the case of off-diagonal disorder.

\begin{figure}
\includegraphics[width=8.4cm]{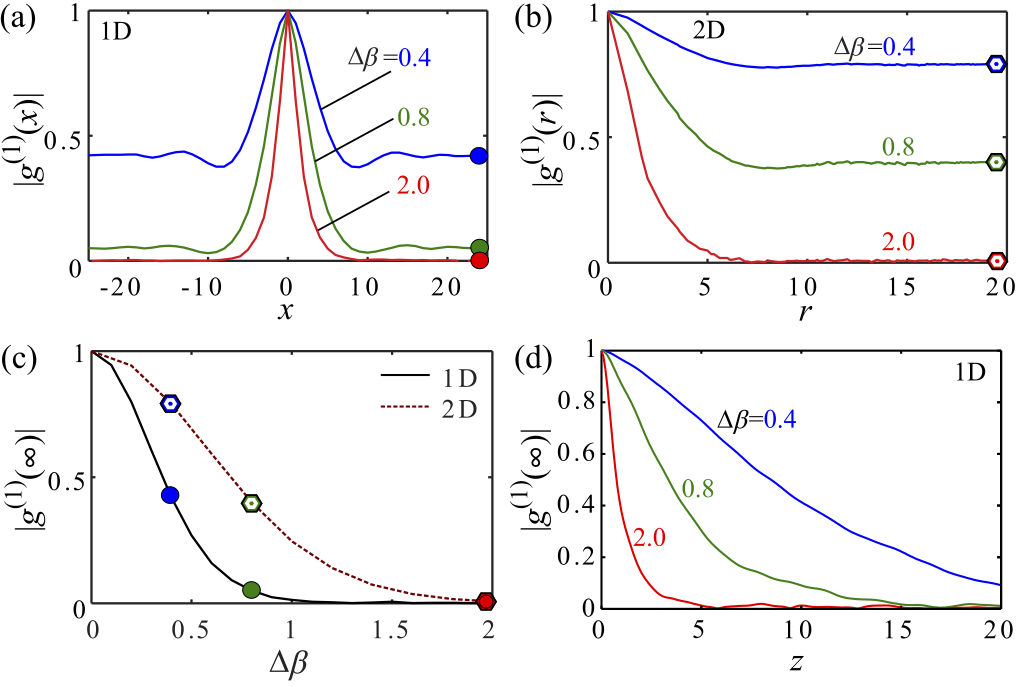}
\caption{\label{figs3} (a) Magnitude of the degree of the normalized transverse coherence function $|g^{(1)}\!(x)|$ for 1D and (b) $|g^{(1)}\!(r)|$ for 2D arrays at various disorder levels at $z\!=\!10$. (c) The long-range-order coherence pedestal $|g^{(1)}\!(\infty)|$ for 1D and 2D arrays. The circles in (a) correspond to those in (c) for 1D lattices. Similarly, the hexagons in (b) correspond to those in (c) for 2D lattices. In (c), 21 and 11 points for $\Delta \beta$  are taken for 1D and 2D arrays, respectively. (d) $|g^{(1)}\!(\infty)|$ as a function of propagation distance $z$ at various disorder levels for 1D waveguide arrays.}
\end{figure}

Next, we calculated the magnitudes of $g^{(1)}\!(x)$ and $g^{(1)}\!(r)$ for 1D and 2D lattices, respectively, revealing a non-zero pedestal $|g^{(1)}\!(\infty)|$ riding on which is a finite-width distribution. The results are presented in Figs.~\ref{figs3}. From these results we obtained the long-range-order pedestal $|g^{(1)}\!(\infty)|$ for 1D and 2D lattices, which are plotted in Fig.~\ref{figs3}(c). Finally, we present the axial evolution of the long-range-order coherence pedestal $|g^{(1)}\!(\infty)|$ for different diagonal disorder levels in a 1D lattice in Fig.~\ref{figs3}(d).

\subsection*{Axial Coherence}		
\begin{figure}
\includegraphics[width=8.4cm]{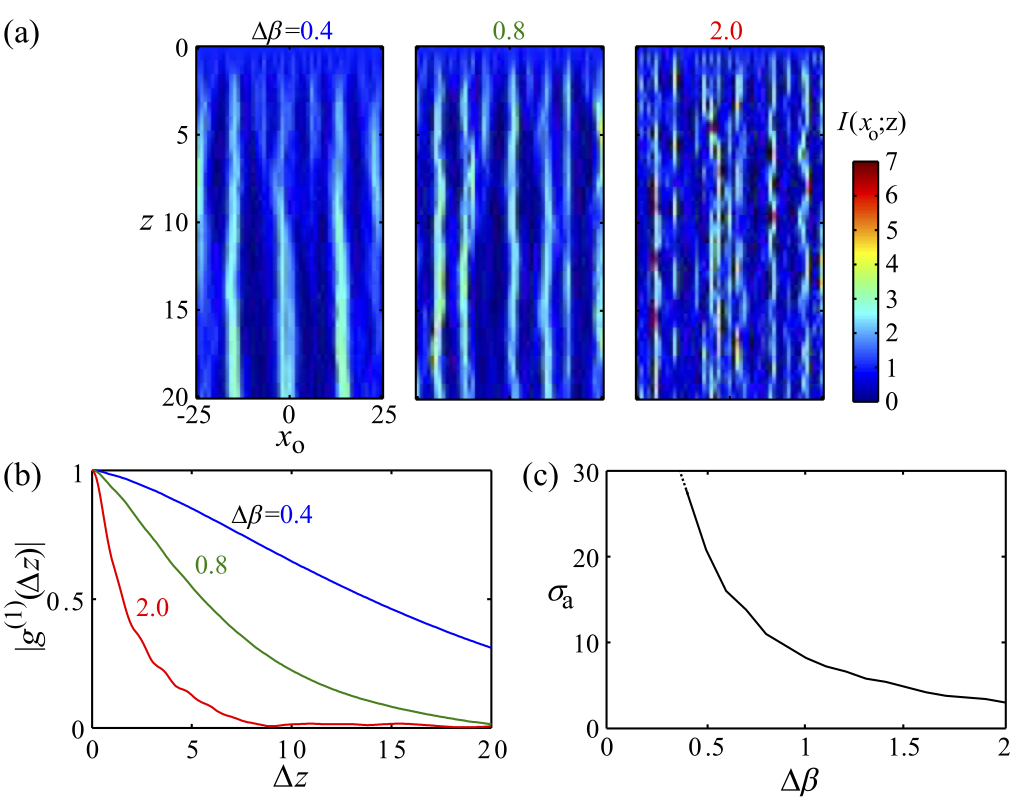}
\caption{\label{figs4}(a) Axial evolution of the intensity along individual realizations of 1D lattice with different disorder levels. (b) The normalized axial coherence function $|g^{(1)}\!(\Delta z)|$ for different disorder levels and (c) its width $\sigma_{\mathrm{a}}$. (a-b) For $z$, 200 points are taken. In (c), 11 points  are taken for $\Delta \beta$.}
\end{figure}

Figure~\ref{figs4}(a) depicts the axial evolution of the intensity along individual realizations of 1D lattices with different disorder levels. The normalized axial coherence function $|g^{(1)}\!(\Delta z)|$ for different disorder levels and its width $\sigma_{\mathrm{a}}$ are given in Fig.~\ref{figs4}(b,c).

\subsection*{Dephasing}		
\begin{figure}
\includegraphics[width=8.4cm]{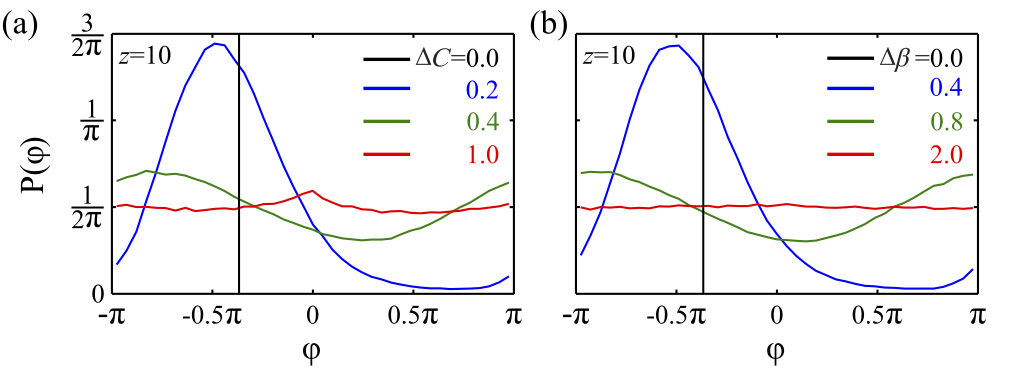}
\caption{\label{fig:phases} Probability distribution function of the phases $\varphi$ of the of the field in (a) off-diagonal  and (b) diagonal 1D disordered arrays at $z\!=\!10$. For periodic arrays, the phase has a constant ($z$-dependent) value (black vertical line at $\varphi\!=\!-0.372\pi$ at $z\!=\!10$). In this case, the distribution is simply a delta function of height 1.}
\end{figure}

\begin{figure}
\includegraphics[width=8.4cm]{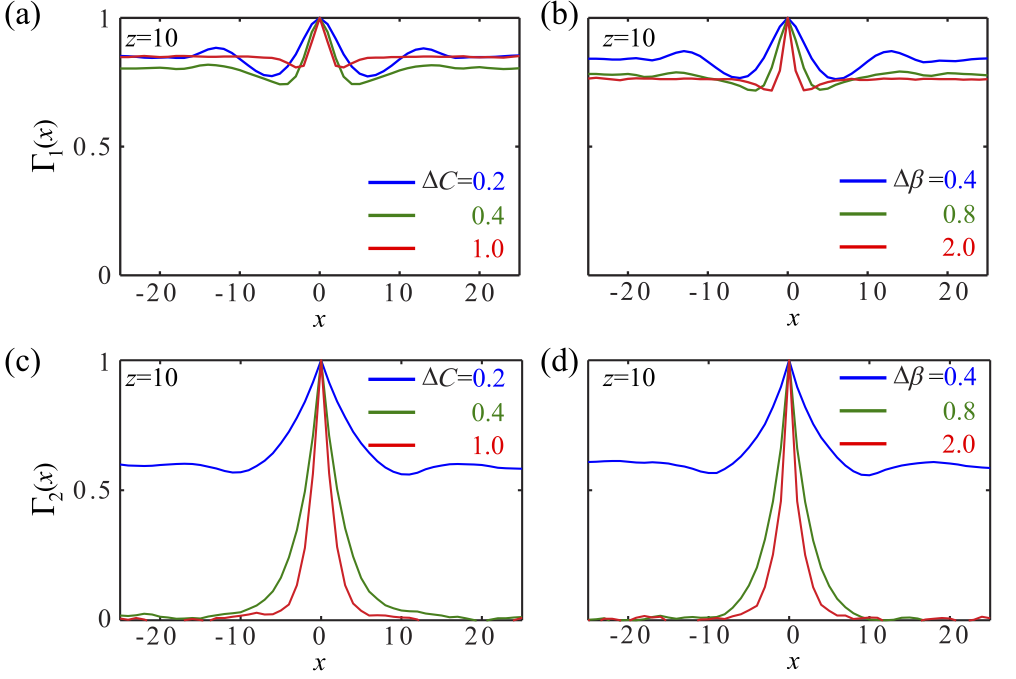}
\caption{\label{fig:PhsAmp1} Transverse coherence function $\Gamma_1(x)\!=\!\langle |E(0)||E^*(x)| \rangle$ for the field magnitude in (a) off-diagonal  and (b) diagonal 1D disordered arrays at $z\!=\!10$. Transverse coherence function $\Gamma_2(x)\!=\!\langle\varphi(0)\varphi(x) \rangle$ for the field phase in (c) off-diagonal  and (d) diagonal 1D disordered arrays at $z\!=\!10$.}
\end{figure}

\begin{figure}
\includegraphics[width=8.4cm]{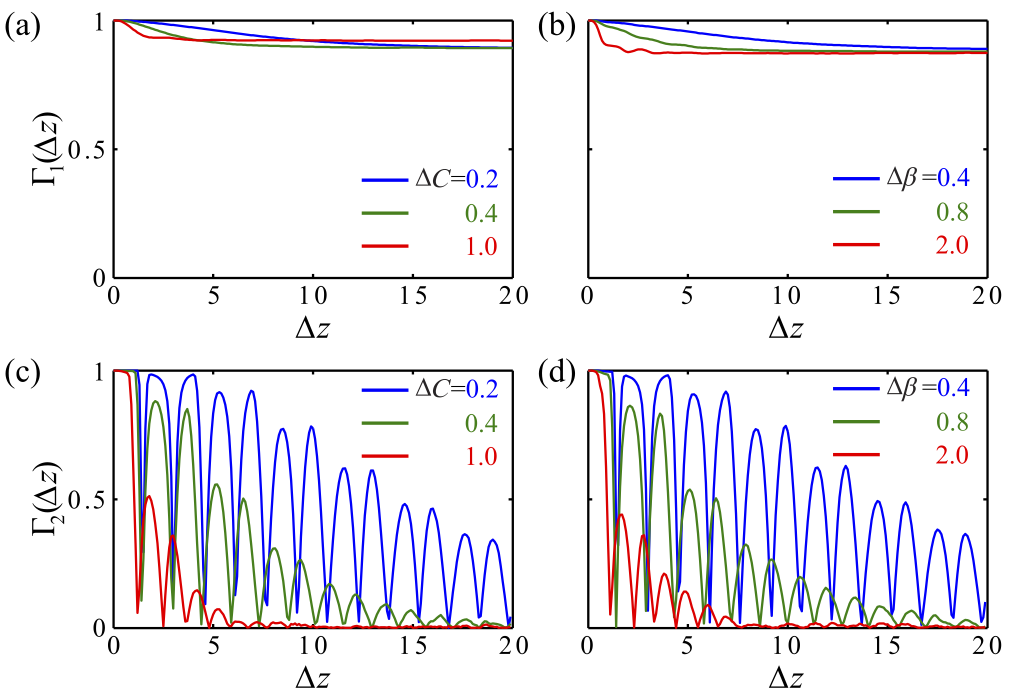}
\caption{\label{fig:PhsAmp2} Axial coherence function $\Gamma_1(\Delta z)\!=\!\langle |E(0)||E^*(z)|\rangle $ for the field magnitude averaged over all transverse positions $x$ in (a) off-diagonal  and (b) diagonal 1D disordered arrays.  (b) Axial coherence function $\Gamma_2(\Delta z)\!=\!\langle\varphi(0)\varphi(\Delta z)\rangle$ for the field phase in (c) off-diagonal  and (d) diagonal 1D disordered arrays also averaged over all transverse positions $x$. }
\end{figure}

Here, we present the probability distributions for the phase of the output field due to extended coherent illumination for both disorder classes in Fig.~\ref{fig:phases}. The probability distribution depends on the propagation distance and disorder level. In all cases, the probability distributions evolve into a uniform distribution as the propagation distance increases. This transition happens quicker for higher disorder levels.  Figure ~\ref{fig:PhsAmp1} depicts the transverse coherence function calculated for the \textit{magnitude} and \textit{phase} of the field separately for both disorder classes. The results confirm that the statistical fluctuations of the field is in fact due primarily to dephasing, not due to fluctuations of the magnitude of the field. Similar behavior can also be observed in the axial coherence function as shown in Fig.~\ref{fig:PhsAmp2}.

\newpage 
\section*{Conclusion}
As we have seen, the qualitative behavior of the field statistical characteristics for 1D and 2D lattices with diagonal disorder is similar to those with off-diagonal disorder. The main quantitative difference that we observed is that we usually double the amount of disorder when going from off-diagonal to diagonal disorder. That is, the quantitative results presented here approach those in the main text most when we set $\Delta\beta=2\Delta C$. We conclude that discrete Anderson speckle produced in waveguide arrays with \textit{diagonal} and \textit{off-diagonal} have the same qualitative statistical characteristics.

%%%%%%%%%%%%%%%%%%%%%%%%%% 

\clearpage 

\vspace*{3cm}

\begin{widetext}

\noindent \Large \textbf{Related articles of the present authors you may be interested in}
\vspace{1cm}

\relatedworkarxiv
{Lattice topology dictates photon statistics}{1611.06662}
{Sci. Rep.}{}{}{}\arxiv{1611.06662}  %\OpenAccess

\relatedworkarxiv
{Interferometric control of the photon-number distribution}{1706.08243}
{APL Photonics}{}{}{}\arxiv{1706.08243}.

\relatedwork
{Hanbury \uppercase{B}rown and \uppercase{T}wiss anticorrelation in disordered photonic lattices}%title
{10.1103/PhysRevA.94.021804}{Phys. Rev. A}{94}{021804(R)}{2016} [\arxiv{1609.07543}]

\relatedwork
{Sub-thermal to super-thermal light statistics from a disordered lattice via deterministic control of excitation symmetry}{10.1364/OPTICA.3.000477}{Optica}{3}{477}{2016} [\arxiv{1609.07562}]

\relatedwork
{A photonic thermalization gap in disordered lattices}%title
{10.1038/nphys3482}{Nat. Phys.}{11}{930}{2015} [\arxiv{1609.07608}]

\end{widetext}

\vfil 

\end{document}